\documentclass[namedreferences,hyperref,optionalrh,solaromanenum]{spr-sola}

\usepackage{graphicx}                    % For eps figures, newer & more powerfull
\usepackage{color}                       % For color text: \color command
\renewcommand{\vec}[1]{{\mathbfit #1}}

\chardef\us=`\_
\usepackage{float}

\usepackage{rotating}

\usepackage{amsmath}

\usepackage{soul}
\usepackage{silence}
\usepackage{multirow}
\usepackage{threeparttable}
\usepackage{rotating}
\usepackage{tikz}

\makeatletter
\newcommand{\undersim}[1]{\mathrel{\mathpalette\@undersim{#1}}}
\newcommand{\@undersim}[2]{%
  \vcenter{%
    \ialign{%
      ##\cr
      $\m@th#1#2$\cr
      \noalign{\nointerlineskip\kern.2ex}
      $\m@th#1\sim$\cr
      \noalign{\kern-.4ex}
    }%
  }%
}

\makeatother

\DeclareRobustCommand{\ion}[2]{%
\relax\ifmmode
\ifx\testbx\f@series
{\mathbf{#1\,\mathsc{#2}}}\else
{\mathrm{#1\,\mathsc{#2}}}\fi
\else\textup{#1\,{\mdseries\textsc{#2}}}%
\fi}

\begin{document}

\begin{frontmatter}

\title{Spectroscopic analysis and RHD modeling of the first Ca II H and H$\epsilon$ flare spectra from DKIST/ViSP}

\author[addressref={1,2,3,5},corref,email={}]{\inits{}\fnm{Cole Tamburri}\snm{}\orcid{0000-0002-3229-1848}}

\author[addressref={1,2,3},corref,email={}]{\inits{}\fnm{Adam Kowalski}\snm{}\orcid{0000-0001-7458-1176}}

\author[addressref={1},corref,email={}]{\inits{}\fnm{Gianna Cauzzi}\snm{}\orcid{0000-0002-6116-7301}}

\author[addressref={1,2,3},corref,email={}]{\inits{}\fnm{Maria Kazachenko}\snm{}\orcid{0000-0001-8975-7605}}

\author[addressref={1},corref,email={}]{\inits{}\fnm{Alexandra Tritschler}\snm{}\orcid{0000-0003-3147-8026}}

\author[addressref={1,2},corref,email={}]{\inits{}\fnm{Rahul Yadav}\snm{}\orcid{0000-0003-4065-0078}}

\author[addressref={1,2},corref,email={}]{\inits{}\fnm{Ryan French}\snm{}\orcid{0000-0001-9726-0738}}

\author[addressref={1,2,3},corref,email={}]{\inits{}\fnm{Yuta Notsu}\snm{}\orcid{0000-0002-0412-0849}}

\author[addressref={1},corref,email={}]{\inits{}\fnm{Kevin Reardon}\snm{}\orcid{0000-0001-8016-0001}}

\author[addressref={4},corref,email={}]{\inits{}\fnm{Isaiah Tristan}\snm{}\orcid{0000-0001-5974-4758}}

\address[id={1}]{National Solar Observatory, University of Colorado Boulder, 3665 Discovery Drive, Boulder, CO 80303, USA}
\address[id={2}]{Laboratory for Atmospheric and Space Physics, University of Colorado Boulder, 3665 Discovery Drive, Boulder, CO 80303, USA}
\address[id={3}]{Department of Astrophysical and Planetary Sciences, University of Colorado Boulder, 2000 Colorado Ave, CO 80305, USA}

\address[id={4}]{Rice University, Houston, TX, USA}
\address[id={5}]{Corresponding Author}

\begin{abstract}
We analyze decay phase observations of the GOES class C6.7 flare SOL2022-08-19T20:31 by the Visible Spectropolarimeter (ViSP) on the National Science Foundation’s Daniel K. Inouye Solar Telescope (DKIST).  The data include the first flare-time DKIST observations of the chromospheric \ion{Ca}{II}~H 396.8 nm and H$\epsilon$ 397.0 nm spectral lines.  These diagnostics have rarely been studied together during the modern era of high-resolution solar flare observations and never at the spectral and spatial resolution of the DKIST.  We directly compare DKIST spectra to state-of-the-art RADYN+RH simulations, including one heated by a non-thermal electron beam and one by in situ thermal conduction.  While certain salient properties of the spectra such as the width of H$\epsilon$ are reproduced, the models severely underestimate the width of Ca II H in the red wing and fail to reproduce the exact relative intensity of Ca II H to H$\epsilon$. The models exhibit a range of chromospheric electron densities contributing to H$\epsilon$ line formation spanning over an order of magnitude.  Unlike the modeled lower-order Balmer-series lines, we find that the width of H$\epsilon$ is not solely related to the high-density upper chromosphere; the widths and intensities are also sensitive to the deeper flare layers.  We outline possible avenues towards improvement of flare models, such as a comprehensive evaluation of flare heating mechanisms in the context of both impulsive and decay phase high-resolution data.

\end{abstract}


\end{frontmatter}

\section{Introduction} \label{sec:intro}
Spatial and spectral evolution of the photosphere, chromosphere, and low corona during solar and stellar flares continues to evade comprehensive and self-consistent modeling (e.g., \cite{fletcher2011,kerr2022r,kerr2023r,kowalski2024}).  According to the CSHKP flare model \citep{carmichael1963,sturrock1966,hirayama1974,kopp_pneuman1976}, flares result from the conversion of magnetic free energy into other forms of energy due to magnetic reconnection.  The interaction between magnetic fields from different topological domains accelerates high-energy particles from the coronal reconnection site along magnetic loops and into the dense chromosphere (e.g., \cite{shibata_magara2011,kontar2017,kowalski2024}).  According to the collisional thick-target model (CTTM, \cite{brown1971,aschwanden2002,holman2011,kowalski2024}), these precipitating particles release most of their energy upon impact in the chromosphere.  This energy release occurs at frequencies across the electromagnetic spectrum; for example, in the ultraviolet and extreme ultraviolet (UV/EUV) 304 \AA\ and 1600 \AA\ emission that we observe as bright flare ``ribbons'' (see e.g. review by \cite{Kazachenko2022r}).  

Past studies have analyzed spectroscopic observations of chromospheric dynamics to deduce the properties of the flaring solar atmosphere.  \cite{graham_cauzzi2015} studied the evolution of emission lines indicating the motion of coronal and chromospheric layers during an X-class flare captured by the Interface Region Imaging Spectrograph \citep[IRIS,][]{depontieu2014}, finding coronal upflows (blueshifts) of up to 300 $\text{km}\;\text{s}^{\text{-1}}$ and chromospheric downflows (redshifts) of 40 $\text{km}\;\text{s}^{\text{-1}}$ in flare-time spectral lines from the observed ribbons.  For the same flare, \cite{graham2020} applied the interpretation by \cite{kowalski2017a} to show that the \ion{Fe}{I}, \ion{Fe}{II}, \ion{Mg}{II}, \ion{C}{I}, and \ion{Si}{II} emission lines can be decomposed into separate Gaussian components revealing the downward motion (``condensation'') of the upper chromosphere with velocities of up to 50 $\text{km}\;\text{s}^{\text{-1}}$ and prominent emission from deeper stationary layers excited by high-energy electrons.  \cite{xu2023} analyzed IRIS observations of an X-class flare with multi-Gaussian fitting of chromospheric and transition region lines and found redshifted components with very large velocities, up to 160 $\text{km}\;\text{s}^{\text{-1}}$. Many additional studies report redshifted, blueshifted, and otherwise asymmetric emission line profiles during flares (e.g., \cite{li2015,tian2015,tian2018,libbrecht2019,french2025_downflows}). 

To reproduce the emission line profiles observed in solar flares and understand the underlying physics, many studies (e.g., \cite{kuridze2015,graham2020,kerr_2024}) compare observations to simulated flare spectra produced using the 1D radiative-hydrodynamic (RHD) code RADYN \citep{carlsson1992,carlsson1995,carlsson1997,abbett1999,carlsson2002,allred2015} and the 1D statistical equilibrium code RH \citep{uitenbroek_2001}.  These models characterize the atmospheric response to a sudden influx of energy from an electron beam injected along a single coronal loop.   With the resulting model spectra, it is possible to constrain atmospheric flare properties such as temperature, electron density, bulk velocity, and microturbulence (e.g., \cite{rutten2003,carlsson2016,kowalski2024}).

Even when employing updated prescriptions for pressure broadening from ions and electrons (the quadratic Stark, or Stark-Lo Surdo, effect\footnote{The quadratic Stark-Lo Surdo effect refers to the splitting of degenerate energy states in atoms without a permanent dipole moment due to the presence of an external electric field (e.g., \cite{kowalski2024}). Johannes Stark and Antonino Lo Surdo studied this effect nearly contemporaneously in 1913 and 1914 (see e.g. \cite{leone2004}).}), there remain discrepancies between the simulated and observed broadening of chromospheric lines. \cite{zhu2019} find that when employing an updated prescription for pressure broadening in RH, an additional multiplicative factor of 30 on the Stark-Lo Surdo width in the simulation source code is needed in order to accurately reproduce the Lorentzian wings of the \ion{Mg}{II} lines (see also \cite{kerr2022r,kerr2023r}). 

Additionally, certain peculiarities of solar flare spectra call into question the source of heating during a flare.  Besides the particle beams discussed above, both thermal conduction from the corona (e.g., \cite{rust1984,polito_2018,ashfield_2021,lorincik_2025}) and heating via Alfv\'en waves (e.g., \cite{fletcher_hudson2008,kerr_2016,reep_2016}) produce distinctly different simulated flare spectra.  When, and in what proportions, these mechanisms are relevant in a realistic flare scenario is not a new question (e.g., \cite{neidig_1989}) but remains a major object of research.

Features observed more often during stellar flares (due in part to the typically higher energies associated with these flares compared to the solar case, and differences in the observed spectral bands) have also raised questions that the finer spatial details in solar data may help to answer.  For a review of this topic, see \cite{kowalski2024}.  Spectra of red dwarf flares show that the hydrogen Balmer lines peak much earlier and become much broader than \ion{Ca}{II} lines (e.g., \cite{hawley_pettersen1991,kowalski2013}), despite the similar chromospheric formation heights of these lines. One example is discussed in \cite{kowalski2024b} using data from a large flare on the dM4.5e star YZ CMi \citep{vida2019} wherein the \ion{Ca}{II}~H 396.8 nm line is much narrower than H$\epsilon$ 397.0 nm (see also \cite{paulson2006}).  Some solar spectra indicate that similar dynamics occur on the Sun when comparing the hydrogen Paschen series to \ion{Ca}{II} IR triplet lines \citep{neidig1984}.  In the context of recent work (e.g., \cite{namekata2020,notsu2025}), flare-time line broadening in solar flares can be used to contextualize similar spectral properties during stellar flares, which are spatially unresolved.

One possible explanation for the difference in line width between \ion{Ca}{II}~H and H$\epsilon$ is that separate heating mechanisms in distinct regions of a flare kernel - namely, a bright impulsive ``core'' and a diffuse gradual ``halo'' \citep{neidig1993,kawate2016,namekata2022} - may be responsible for emission in different lines.  The Balmer series emission lines, which are highly sensitive to electric pressure broadening from the linear Stark-Lo Surdo effect (distinct from the aforementioned quadratic Stark-Lo Surdo effect; see e.g. \cite{johnskrull1997,allred2006,namekata2020,kowalski2022a}) and therefore have widths with a strong dependence on the electron density $n_e$, may be produced in the high density condensation or other chromospheric layers due to heating from a high flux electron beam in the kernel core.  The \ion{Ca}{II} lines, which are less broad and evolve more gradually, may be additionally excited in the kernel halo by radiative back-warming from the cores \citep{kowalski2022b,namekata2022b}.  

Although flares are prevalent on both our Sun and other stars, only for the Sun do we have access to direct, high-spatial-resolution observations.  With these, we can capture the evolution of flare spectral lines while also tracking the spatial evolution of the plasma in the photosphere, chromosphere, and corona.  Unfortunately, solar data have often lacked the necessary spectral coverage to provide an unambiguous explanation for differences in the width of the broad hydrogen Balmer series and narrow \ion{Ca}{II} lines.  While there have been solar observations of, for example, the \ion{Ca}{II}~H and H$\epsilon$ lines in the past (e.g., \cite{Svestka1976_rcc}), these lines have not been observed at high spectral resolution in the modern era of high-cadence, high-resolution observations.

Several instruments on the newly commissioned Daniel K. Inouye Solar Telescope (DKIST, \cite{rimmele2020}) are now able to provide spectroscopic and spectropolarimetric observations of flaring regions in multiple spectral windows. In particular, the Visible Imaging Spectropolarimeter (ViSP, \cite{dewijn2022}) is poised to provide novel insight on the physical mechanisms producing chromospheric flare emission lines by producing spatially and spectrally resolved data.  For a more detailed discussion of questions regarding line broadening that DKIST data may help to answer, see \cite{kowalski2022a}.

In this work, we present DKIST observations of a GOES X-ray class C6.7 flare at 20:42 UT on 19 August 2022, including the \ion{Ca}{II}~H 396.8 nm and H$\epsilon$ 397.0 nm line.  While observing conditions were not optimal, these are the first flare observations of these lines with DKIST, representing a unique opportunity to assess the diagnostic power of these lines in combination with modern flare simulations. We perform an initial comparison of DKIST spectra in this spectral range to simulated spectra from the RADYN and RH codes.  

In Section \ref{sec:summary} we describe flare observations collected with DKIST.  In Section \ref{sec:19aug_line_ev} we detail the evolution of \ion{Ca}{II}~H and H$\epsilon$ during the flare and describe differences in \ion{Ca}{II}~H emission line profiles across the ribbon.  In Section \ref{sec:results} we compare data-inspired RHD models of these lines to the observed emission lines.  In Section \ref{sec:disc} we discuss our findings and identify potential paths towards improved modeling of these chromospheric lines.  We summarize our work in Section \ref{sec:conc}.

\section{Summary of Observations}\label{sec:summary}

Observations with the ViSP and Visible Broadband Imager (VBI, \cite{woger2021}) at DKIST were obtained for this experiment between 15 August 2022 and 25 August 2022.  Data for all observations are publicly available in the DKIST data center archive. Table \ref{tab:summobs} summarizes the dates and times during which observations were taken with a description of observed features for the entirety of the experiment.  Of particular interest to this work, on 19 August 2022 we observed the gradual decay phase of the GOES class C6.7 solar flare SOL2022-08-19T20:31 originating in active region AR13078 located at roughly 31S, 59W ($\mu =$ 0.48) from 20:42 UT to 20:46 UT.  ViSP was also observing from 20:55 to 21:16 UT, when there was still some emission in \ion{Ca}{II}~H, though significantly less than at 20:42 UT. Seeing conditions were poor during this window, and VBI-red was not observing.  In our work we focus on the first DKIST observations taken during the C6.7-class flare.

\begin{table}

\caption{Summary of observations}\label{tab:summobs}
\resizebox{\textwidth}{!}{
\begin{tabular}{|c|c|c|c|c|c|p{4cm}|}
\hline
\textbf{Date} & \textbf{Start (UT)} & \textbf{End (UT)} & \textbf{Observed Feature} & \textbf{Fig. 1 Ref.} & \textbf{Data Center Product ID}
\\
\hline
August 15 & 18:00:24 & 19:17:41 & Sunspot, granulation, filament& N/A & L1-OOHKH, L1-IIESO, L1-DJLQE, L1-MQRQD, L1-PFDPU\\
August 15 & 22:02:20 & 23:00:36 & Sunspot, filament & N/A & L1-HTKTO, L1-CEDDM, L1-EIPBQ, L1-TDAQC, L1-YVJHF  \\
August 16 & 17:21:56 & 19:48:25 & Sunspot, granulation, filament & N/A & L1-NNEBO, L1-TTYCV, L1-BAVNX, L1-GPTZR, L1-QSPMJ\\
August 19 & 18:06:59 & 20:04:19 & Pre-flare sunspot & Blue & L1-ZRCBA, L1-WOBUV, L1-KMAZT\\
\textbf{August 19} & \textbf{20:42:07} & \textbf{20:46:50} & \textbf{C6.7 flare, sunspot, filament} & \textbf{Magenta} & \textbf{L1-TDMFC, L1-MQEFD, L1-GPCAR,L1-BAHBX, L1-PFQYU}\\
August 19 & 20:55:55 & 21:16:22 & Ca II H in emission & N/A & L1-ALRBG, L1-NZXQB, L1-DJPJE\\
August 19 & 22:34:22 & 23:23:39 & Sunspot, filament & Yellow & L1-HTTCO, L1-CEEEM, L1-OCZBK, L1-UWJDY, L1-IBWPS\\
August 25 & 17:36:10 & 20:28:42 & Sunspot, granulation, filament & N/A & L1-VKOSH, L1-WOHQV, L1-LYDIP, L1-RUMIL, L1-JNYVI \\ 
\hline

\end{tabular}
}
{\raggedright \tiny \textit{Note. Summary of observations for PID 1.84 between 2022 August 15 and 2022 August 25, including start and end times, observed features, and references to relevant time periods in Figure \ref{fig:goes_summary}(a).  The observations that are used for analysis of flare lines in this work are highlighted in bold.  DKIST data center identifiers are listed in the last column; see https://dkist.data.nso.edu/.}\par}

\end{table}

\subsection{Instrument configuration}\label{sec:inst_config}

ViSP spectral coverage included (i) the \ion{Fe}{I} 630.2 nm line on arm 1, with a spectral range of ${\sim}1$~nm and (ii) the \ion{Ca}{II}~H 396.85 nm and H$\epsilon$ 397.01 nm lines together on arm 2, with a spectral range of ${\sim}$0.76 nm.  Although we are interested only in the intensity profiles of the lines observed, the option for collecting only Stokes I observations was unavailable during DKIST Observing Cycle 1.  Therefore, the instrument was observing in polarimetric mode at a frame rate of 33 Hz with 10 modulation states and 2 modulation cycles.  The slit width was 0.1071'', and ViSP performed a small four-step raster scan with a step size of 1.99''. The slit lengths of Arm 1 (\ion{Fe}{I} 630.2 nm) and Arm 2 (\ion{Ca}{II}~H 396.8 nm) are 75''.8 with spatial sampling 0''.0298 $\text{pix}^{\text{-1}}$ and 62''.3 with spatial sampling 0''.0245 $\text{pix}^{\text{-1}}$ respectively.  The time between steps was 1.8 s for a scan time of ${\sim}$7.2 s, though instrument overhead lengthened the interval between successive map repeats (the map cadence) to 26 s.  The scan was repeated ten times.  After wavelength calibration (Appendix \ref{sec:app1}), we find the spectral sampling of the \ion{Fe}{I} and \ion{Ca}{II}~H arms to be 12.7 m\AA\ and 7.7 m\AA\ respectively.  

We determine the achieved spatial resolution of the data along the slit by examining the spatial frequency at which there is no significant power from solar structures; higher frequencies are dominated by instrumental or white noise. We find that this occurs for spatial scales smaller than ${\sim}$1'' $\approx$ 720 km. This was due to poor seeing conditions (the median Fried parameter was $r_0 =$ 3.7 $\pm$ 0.9 cm during the flare observing period).

VBI data contextualize the ribbon dynamics observed in the ViSP spectra. VBI-red (69'' x 69''\ FoV) observed in (i) offband H$\alpha$ 656.3 nm (effective filter width $\sim\text{0.65}\;\text{\AA}$) and (ii) TiO 705.8 nm filters (filter width $\sim\text{6}\;\text{\AA}$).  VBI-blue (45'' x 45''\ FoV) collected data from a filter centered in the continuum at 450.3 nm (filter width $\sim\text{4}\;\text{\AA}$).  The VBI data were speckle-reconstructed \citep{woger2008}, with each reconstructed image produced from 80 raw images.  The individual raw images have exposure times of 0.07 ms, 1 ms, and 7 ms for the TiO, H$\alpha$, and blue continuum bands respectively. Speckle-reconstructed data from VBI-red and VBI-blue have cadence of 6.1 and 2.7 s and spatial sampling 0''.017 and 0''.011 pix$^{-1}$, respectively.

\subsection{19 August 2022 SOL2022-08-19T20:31 flare}\label{flaresum}

\begin{figure}
\includegraphics[width=\textwidth]{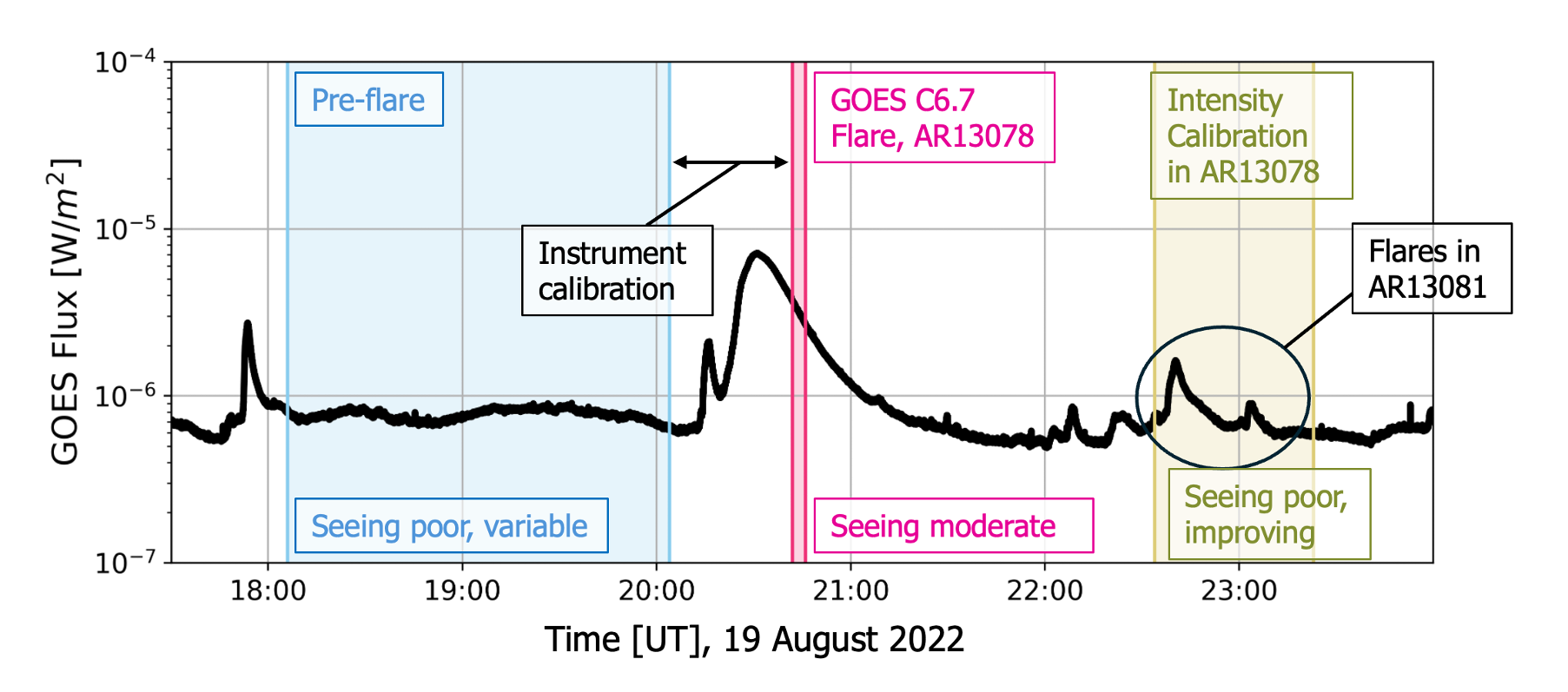}
\centering
\caption{\tiny Overview of GOES/XRS data during flare observations. We show the light curve in GOES/XRS 1-8 \AA{} SXR (black) with shaded regions indicating the periods when DKIST was observing.  DKIST was observing exclusively during the periods indicated by colored shading (blue, red, yellow).  The red shaded region indicates the period when DKIST was observing at flare-time.  DKIST was undergoing instrument calibrations between the blue pre-flare period and the flare observations.  The data in the final, yellow-shaded window are used for intensity calibration in this work.  The two flares observed by GOES during this period occurred in a different active region, AR13081.}
\label{fig:goes_summary}
\end{figure}

\begin{figure}
\includegraphics[width=.8\linewidth]{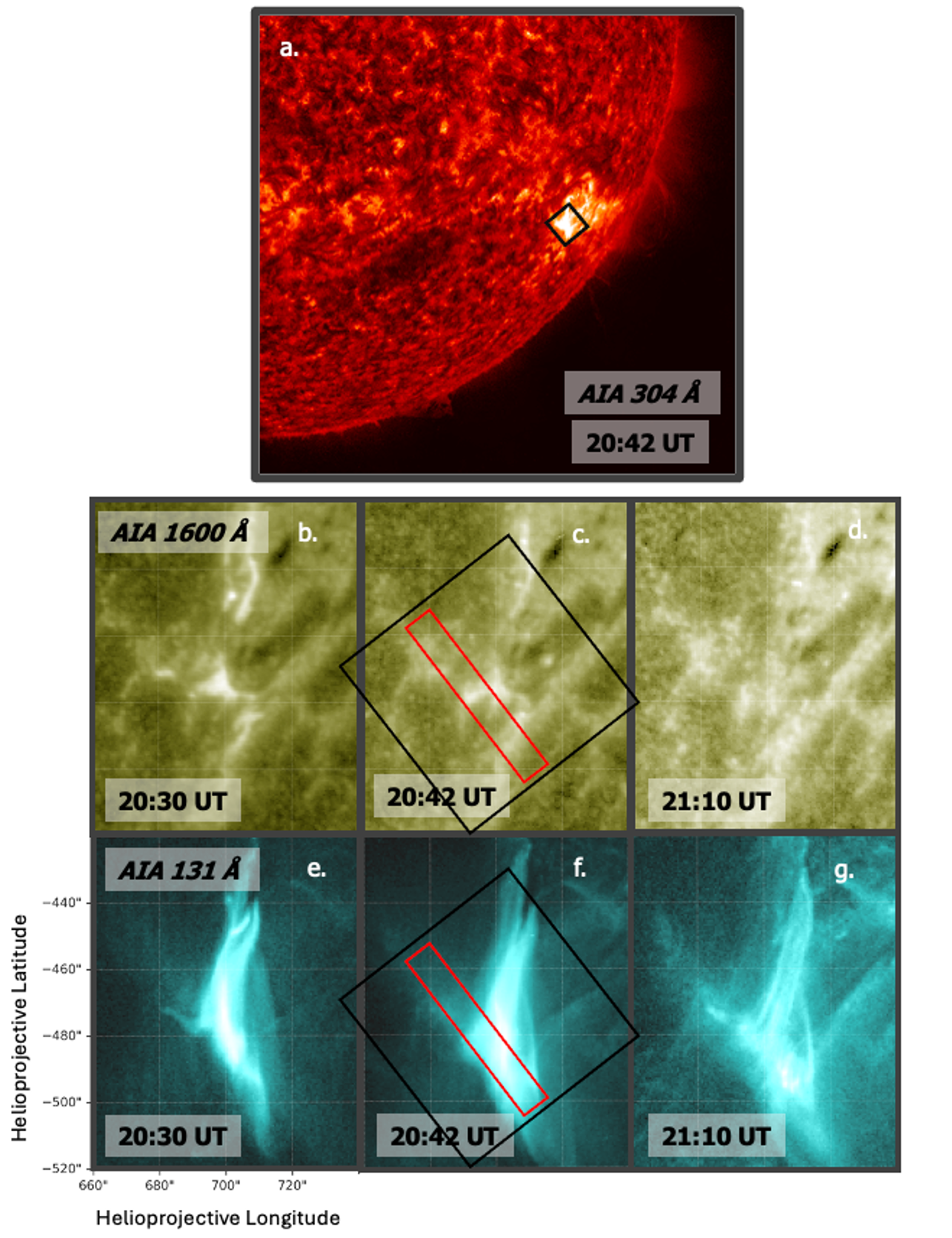}
\centering
\caption{\tiny SDO/AIA images from 2022 August 19 for three time-steps before (20:30 UT, panels (b) and (e)), during (20:42 UT, panels (a), (c), (f)), and after (21:10 UT, panels (d) and (g)) DKIST flare observations.  Panel (a), (b) through (d), and (e) through (g) show the 304 \AA, 1600 \AA, and 131 \AA{} images respectively.  Panels (a), (c), and (f) correspond to the beginning of flare observations with DKIST.  The approximate FoV of VBI (black) and the ViSP raster scan (red) are shown in panels (a), (c), and (f).  The ViSP FoV intersects with the southern flare ribbon as seen in 1600 \AA{} images.  The coronal loops corresponding to the ribbons seen in the 1600 \AA{} images are visible in the 131 \AA{} images in panels (e)-(g).}
 \label{sdocontext}
\end{figure}

\begin{figure}
\includegraphics[width=.9\linewidth]{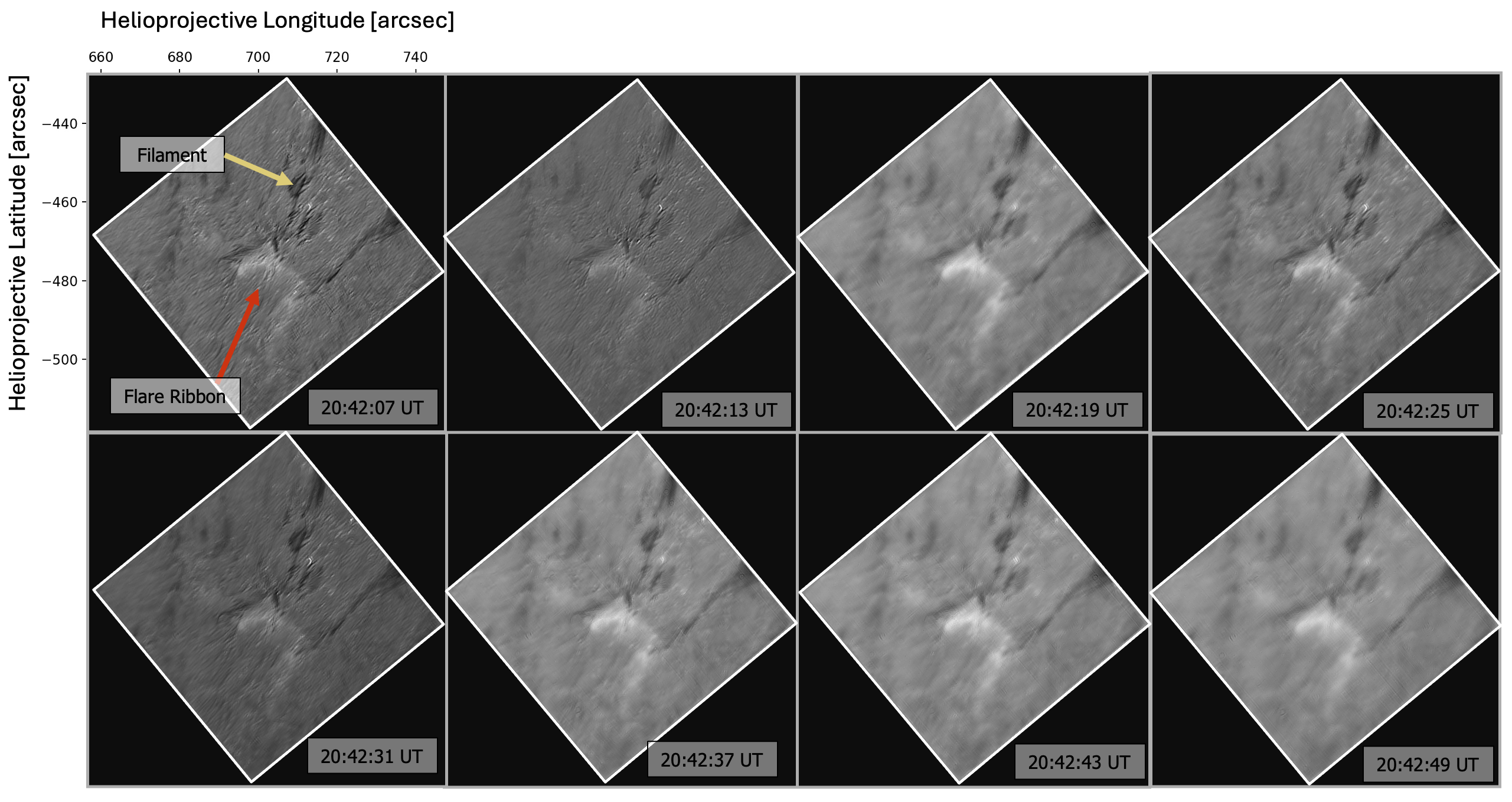}
\centering
\caption{\tiny VBI H$\alpha$ observations of the GOES C6.7 class flare in NOAA active region 13078 on 2022 August 19. The bright flare ribbon and dark filament structures, used for co-alignment with ViSP spectra, are clearly visible, particularly in the first image frame (upper left).  There are significant variations in image quality.  The VBI-red field of view is 69''\ x 69''. Helioprojective latitude and longitude values refer to those determined from co-alignment with SDO (Appendix \ref{sec:app2}).} 
\label{VBI_summary}
\end{figure}

The magenta shaded region of Figure \ref{fig:goes_summary}(a) shows the period corresponding to the 19 August 2022 C6.7-class flare data from DKIST/ViSP overlaid on a plot of the GOES SXR 1-8 \AA\ flux.  A subset of the data taken during the yellow shaded period in Figure \ref{fig:goes_summary} beginning at 22:34 UT best represents the non-flaring Sun for this dataset (the flares observed in the GOES light curve during this period occurred in a different region, AR13081).  We use a spatially and temporally averaged spectrum from this period as the ``non-flare'' spectrum throughout.  We compare the non-flare data to the Neckel solar atlas spectra \citep{neckel1984,neckel1999} to perform intensity calibration for flare-time spectra, discussed in Appendix \ref{sec:app1}.  We define rest wavelengths $\lambda_c$ by comparison to the quiet-Sun, disk-center atlas after wavelength calibration.  Figure \ref{sdocontext} shows the VBI and ViSP fields-of-view overlaid on three Solar Dynamics Observatory Atmospheric Imaging Assembly (SDO/AIA, \cite{lemen2011}) channels during the times corresponding to the flare.  Panels (b), (c), and (d) in Figure \ref{sdocontext} show the evolution of the flare ribbons in the 1600 \AA\ channel, with the ViSP and VBI fields-of-view overlaid.    Figure \ref{VBI_summary} shows the first eight VBI-red H$\alpha$ images of the flare.  Reported helioprojective latitude and longitude coordinates are determined by co-alignment with SDO/AIA (Appendix \ref{sec:app2}).

\begin{figure}
\includegraphics[width=\linewidth]{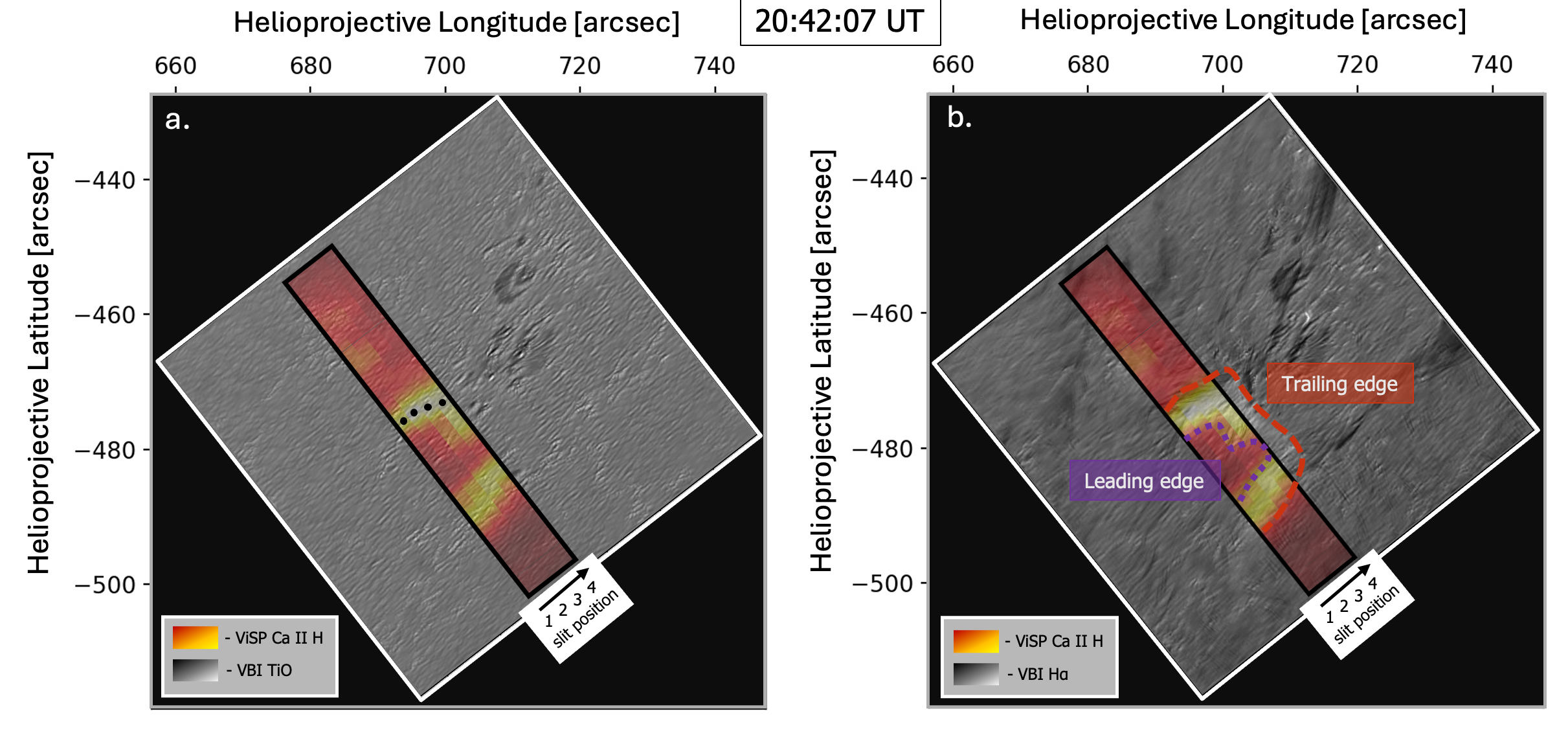}
\centering
\caption{\tiny Result of co-aligning DKIST/ViSP \ion{Ca}{II}~H 396.8 nm integrated intensity maps with DKIST/VBI TiO and H$\alpha$ images from the first ViSP scan and first VBI image and determining proper spatial coordinates by comparison to SDO, as discussed in Appendix \ref{sec:app2}. (a) The result of transforming VBI image and ViSP integrated intensity map into the SDO basis, with the ViSP \ion{Ca}{II}~H integrated intensity from the first scan at 20:42:07 UT overlaid on a VBI TiO filter image.  (b) The same as panel (a), but with the \ion{Ca}{II}~H integrated intensity overlaid onto the VBI-red H$\alpha$ filter image at 20:42:07 UT.  We indicate the location of the ribbon leading and trailing edges (defined in Section \ref{sec:CaIIvary}).  In both panels we indicate the direction of the ViSP scan, with the four slit positions corresponding to panels (a)-(d) in Figure \ref{fig:flare_summary_19_august} respectively.  The raster step is 1.99''\ and the width of the slit is enlarged for clarity.  Black dots in (a) correspond to the positions of maximum intensity, corresponding to the earliest intensity profiles in Figure \ref{fig:flare_summary_19_august}.}
 \label{context}
\end{figure}

In Figure \ref{context} we overlay VBI-red TiO and H$\alpha$ images with the ViSP observations integrated across the \ion{Ca}{II}~H 396.8 nm line. There are two regions of elevated emission.  By comparison to SDO/AIA as shown in Figure \ref{sdocontext}, we find that the brightening nearest to the center of the ViSP slit corresponds to the location of a flare ribbon rooted at the southern footpoint of an arcade of coronal field loops as identified in the 131 \AA\ channel of SDO/AIA (Figure \ref{sdocontext} (e), (f), and (g)). The southern, less bright feature captured by ViSP (evident in Figure \ref{context}(a)) is part of the same ribbon, which is geometrically curvilinear and visible in its entirety in Figure \ref{sdocontext}(c).  The edges of the ribbon that ``lead'' and ``trail'' the motion of the main ribbon are indicated in Figure \ref{context}(b).  Judging from co-alignment with VBI and SDO/AIA in Figure \ref{context}, the ViSP slit lies roughly perpendicular to the regions of elevated flare intensity observed in the ribbon.  The conjugate northward flare ribbon (visible in Figures \ref{sdocontext}(b) and (c)) was not captured by ViSP. The small portion of the northern ribbon that may be within the VBI FoV is apparently obscured by an overlying dark filament in the H$\alpha$ images.  For initial analysis of line evolution in Figure \ref{fig:flare_summary_19_august}, we consider the pixels corresponding to the location of highest intensity for each position in the raster (black dots in Figure \ref{context}(a)).

\section{Spectroscopic Analysis of ViSP Data}\label{sec:19aug_line_ev}

\subsection{Summary of observed Ca II H and H\texorpdfstring{$\epsilon$}{e} emission lines}

\begin{figure}
\includegraphics[width=\linewidth]{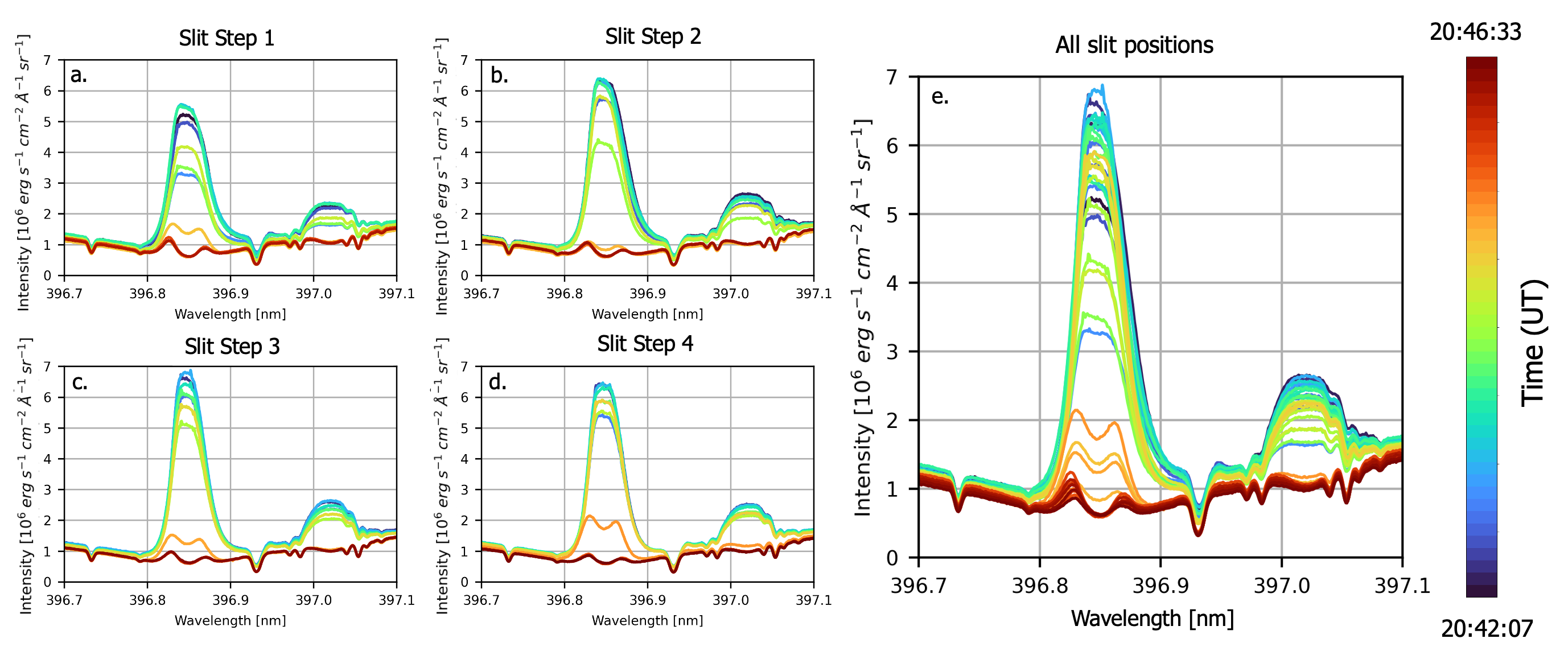}
\centering
\caption{\tiny Intensity-calibrated ViSP spectra.  In panels (a)-(d) we show \ion{Ca}{II}~H 396.8 nm and H$\epsilon$ 397.0 nm emission line profiles for individual slit positions per scan.  Spectral profiles are taken at the position of maximum intensity in the brightest flare ribbon.  In panel (e) we combine all spectra from panels (a)-(d) in the observations between 20:42:07~UT and 20:46:33~UT (red shaded region in Figure \ref{fig:goes_summary}(a)).  For the first scan at 20:42:07UT, the spectra correspond to the positions marked by the black dots in Figure \ref{context}(a).}
\label{fig:flare_summary_19_august}
\end{figure}

As described in Section \ref{sec:inst_config}, the ViSP was observing in polarimetric mode, though in this work we investigate only the evolution of Stokes I profiles.  Figure \ref{fig:flare_summary_19_august} shows the emission line profiles of \ion{Ca}{II}~H and H$\epsilon$ at the flare ribbon center for the entirety of DKIST observations from 20:42:07 UT to 20:46:33 UT. Both \ion{Ca}{II}~H and H$\epsilon$ were in emission at the start of observations, but while \ion{Ca}{II}~H was in emission throughout the observations, H$\epsilon$ returns to the non-flare profile at 20:44:43 UT and is not appreciably in emission thereafter.  At the beginning of the observing period, \ion{Ca}{II}~H is single-peaked with peak maximum intensity slightly in the blue and a broader red than blue wing at the brightest ribbon location.  \ion{Ca}{II}~H becomes less intense throughout the observations, but the decrease in intensity is not constant, likely due to variations in observing conditions.  At the location of maximum intensity in the ribbon, \ion{Ca}{II}~H develops a double-peaked profile with a blue peak maximum at 20:45~UT and remains double-peaked thereafter.  While in emission, H$\epsilon$ is symmetric and single-peaked.  By the end of the ViSP observations at 20:46 UT, H$\epsilon$ has effectively returned to pre-flare levels.  However, emission in \ion{Ca}{II}~H was still observed by ViSP at 20:55 UT (not shown), though this dataset lacks context images from VBI-red. 

Since the map cadence is low and very little variation in line shapes is observed, we primarily describe spatial variations in spectral line shape during the first step of the first ViSP scan.

\subsection{Spatial variations in Ca II H \textbf{and H\texorpdfstring{$\epsilon$}{e} line asymmetry}}\label{sec:CaIIvary}

\begin{sidewaysfigure}
\includegraphics[width=1\linewidth]{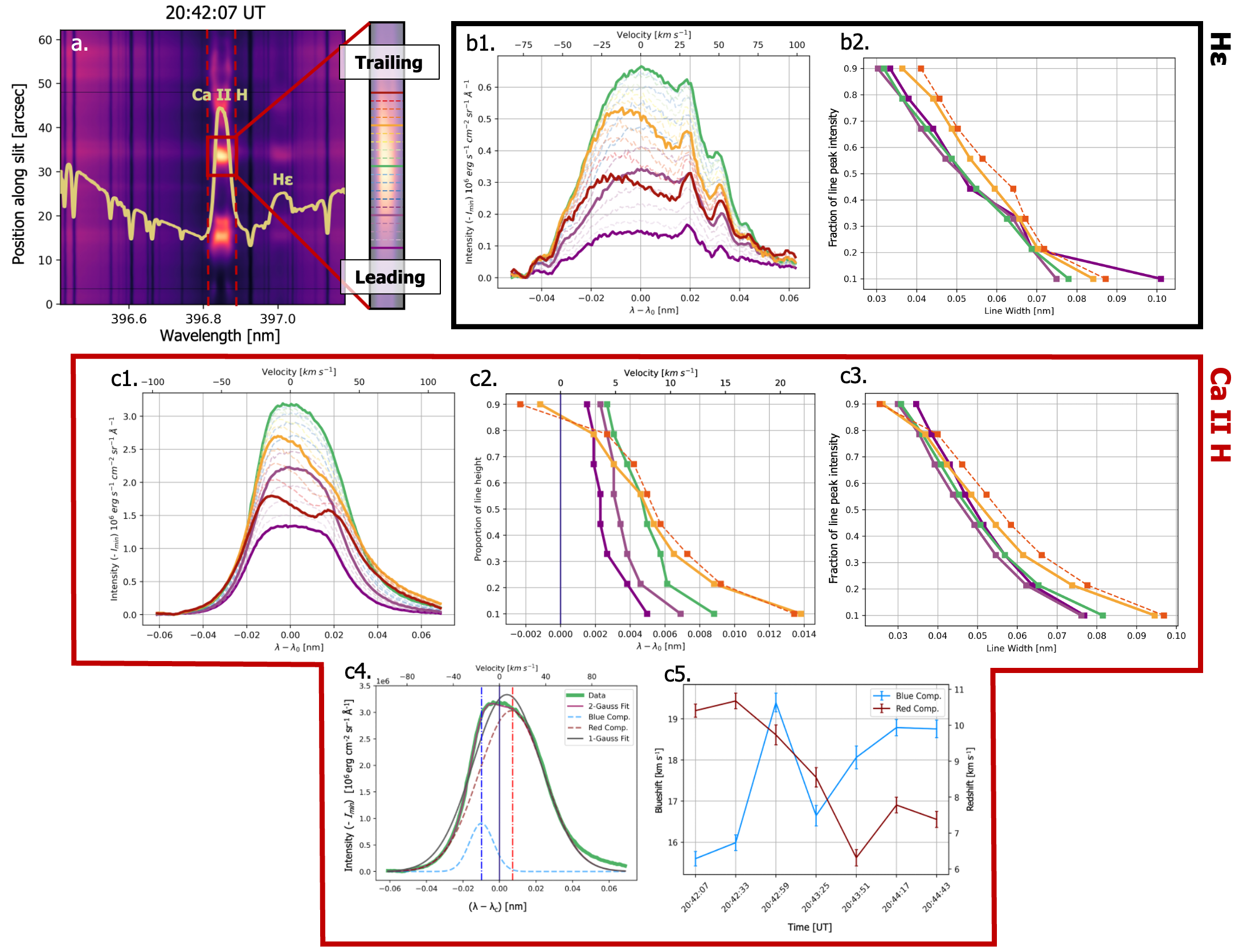}
\centering

\caption{\tiny Summary of the \ion{Ca}{II}~H and H$\epsilon$ emission lines across a flare ribbon.  (a) Calibrated slit spectrum before subtraction of non-flare spectrum.  We indicate the ribbon of largest intensity captured by the ViSP within the red box.  The ViSP spectrum at the position of maximum ribbon intensity is shown in yellow (corresponding to the location of the green horizontal line in the cut-out image).  Horizontal lines in the cut-out image indicate the slit positions corresponding to the spectra in b1 and c1.  Solid dark violet, light violet, green, orange, and red horizontal lines correspond to the leading edge, inner leading edge, ribbon peak intensity, inner trailing edge, and trailing edge of the ribbon.  (b1) Background-subtracted H$\epsilon$ line profiles at the horizontal lines in (a).  Peaks in the red wing are due to the presence of \ion{Fe}{I} absorption lines. (b2) H$\epsilon$ line widths for the profiles in b1.  Due to the presence of the \ion{Fe}{I} absorption lines, these are calculated by doubling the width in the blue wing of H$\epsilon$. (c1)  Background-subtracted \ion{Ca}{II}~H emission lines corresponding to horizontal lines in (a).  (c2) Bisector positions for the leading edge, inner leading edge, ribbon center, inner trailing edge, and near trailing edge profiles with the same colors and line-styles as (a). We do not include the true trailing edge; the profile there is double-peaked, so a bisector can not be calculated near $\lambda_c$.  (c3) Line widths for the same profiles as are analyzed in (c2). (c4) Example double-Gaussian decomposition (solid violet line) of the ribbon peak intensity emission line (green line) with red and blue Gaussian components (dashed dark red and dashed blue lines, respectively).  The best-fit single Gaussian is over-plotted in grey for comparison.  (c5) The evolution of blue and red Gaussian component velocity shift from line center in the first slit position. Wavelengths are reported relative to the rest wavelength for \ion{Ca}{II}~H line center at $\lambda_c = 396.85$ nm.}
\label{kernzoom}
\end{sidewaysfigure}

Given the very few positions in each ViSP scan and $\sim$1''\ spatial resolution, it is not possible to resolve fine-structure ribbon features such as the ribbon front (e.g. \citealt{sharykin2014,jing2016,polito_2023,kerr_2024}). Still, there are clear differences in spectral line shape across the ribbon. In Figure \ref{kernzoom}(a) we show the spectral line intensity from the first scan of the ViSP.  We display the line profiles of H$\epsilon$ and Ca II H across the brightest flare ribbon located near the center of the slit (Figure \ref{kernzoom}(b1) and (c1)), noting the profiles at the ribbon leading edge (violet), inner leading edge (light violet), ribbon center (green), inner trailing edge (orange) and trailing edge (red).  The ``inner" leading and trailing ribbon edges are defined to bound the region for which the \ion{Ca}{II}~H profile is clearly single-peaked rather than flat-topped (closer to the leading edge) or double-peaked (nearer the trailing edge).  Near the leading edge of the ribbon (violet), the emission profile is first flat-topped and fairly symmetric, shifted to the red.  Closer to the ribbon center (green), the profile becomes more asymmetric with a broader red wing.  At ribbon center and towards the trailing edge of the ribbon (orange), the ribbon is asymmetric with a blue peak.  The asymmetry becomes more significant closer to the trailing edge (dark red), where we note the presence of a central reversal (or double-peaked line profile).  H$\epsilon$ demonstrates these same features, though less clearly due to the influence of the Fe I lines in the red wing.

We quantify the Ca II H asymmetry observed in the ribbon center emission line profile using three methods.  We discuss H$\epsilon$ to a lesser degree because the presence of \ion{Fe}{I} absorption lines in the red wing makes a definitive determination of H$\epsilon$ line widths difficult.  We adjust for viewing angle $\mu$ in the quantification of the lineshifts.  First, in Figure \ref{kernzoom}(c2) we show the variation in the bisector position at increments of 10\% of peak intensity for each of the profiles.  This method is often used to extract mass motion velocities from flare line profiles without the use of computationally expensive simulations \citep{canfield1990,ding1995,graham_cauzzi2015,graham2020}.  At each tenth percentile in the emission line profile, we determine the wavelength halfway between the red and blue wings at that intensity value.  One possible interpretation of this analysis is that the relative position of the bisector indicates whether there is upward or downward mass motion (an ``evaporation-condensation'' model) at the heights probed by the emission line.  Following e.g. \cite{graham_cauzzi2015}, the 30\% bisector velocity could be interpreted as a lower limit on the condensation velocity of the line at ribbon center, corresponding to $v_{cond} \approx \text{4.35}\;\text{km}\;\text{s}^{\text{-1}}$.  However, without detailed modeling, we can not be certain that this value indicates the mass motion direction (e.g., \cite{heinzel1994,kuridze2015}). 

Comparison of the bisectors in Figure \ref{kernzoom}(c2) illustrates the difference in red- and blue-shifted components across the flare ribbon.  The blue vertical line indicates the line center of \ion{Ca}{II}~H.  At the leading edge (violet), there is a greater redward asymmetry in the lower percentile levels, while the upper levels are shifted less so in the red.  The redward asymmetry at lower bisectors becomes more significant closer to the ribbon center.  While the redward asymmetry continues to increase, there arises a stronger blueward asymmetry towards the trailing edge (red).  Both patterns - the increase in the redward asymmetry in lower percentiles and of the blueward asymmetry in the upper percentiles - are notably monotonic.  

In Figure \ref{kernzoom}(c3) we show the line width of \ion{Ca}{II}~H at the bisector positions in panel (c2).  Generally, \ion{Ca}{II}~H is wider at the trailing edge than the leading edge or ribbon center for bisector positions below 70\%\ of the line peak, particularly in the far wing.  The line width at the 30\% bisector position is ${\sim}$0.01 nm larger at the trailing edge than at the leading edge (${\sim}$0.065 nm vs. ${\sim}$0.055 nm). In panel (b2) we similarly show the H$\epsilon$ line width, though in this case we define the line width as twice the width of the blue wing relative to the central wavelength.  This is done to avoid any influence from the lines in the red wing.  The H$\epsilon$ line width follows a pattern across the ribbon similar to the Ca II H line width.

Lastly, we perform a Gaussian decomposition of the observed single-peaked \ion{Ca}{II}~H profiles, similar to the double- or multi-Gaussian models employed for other lines (e.g. \cite{graham2020} and \cite{xu2023}).  An example is shown in Figure \ref{kernzoom}(c4).  Using the python package \texttt{scipy.optimize.leastsq}, we fit the \ion{Ca}{II}~H emission lines in each image frame to a double-Gaussian model.  The \ion{Ca}{II}~H line profile at ribbon center is well-represented by a double Gaussian composed of a blueward component and a redward component centered at $v_{blue} = -\text{15.6}\;\text{km}\;\text{s}^{\text{-1}}$ and $v_{red} = \text{10.4}\;\text{km}\;\text{s}^{\text{-1}}$ respectively.  Uncertainties in the fit results are derived from the covariance matrix output of \texttt{scipy.optimize.leastsq}, which uses the Levenberg-Marquardt optimization algorithm. Figure \ref{kernzoom}(c5) shows the evolution of the blue and red Gaussian component central wavelengths, expressed as velocity shifts, for the full evolution of the \ion{Ca}{II}~H profile at ribbon center in the first step of each scan.   There is a slight increase in blueshifted component velocity (${\sim}\text{16}\;\text{km}\;\text{s}^{\text{-1}}$ to ${\sim}\text{19}\mathrm{\;km\;s^{-1}}$) and decrease in redshifted component velocity (${\sim}\text{10}\;\text{km}\;\text{s}^{\text{-1}}$ to ${\sim}\text{7}\;\text{km}\;\text{s}^{\text{-1}}$) from 20:42:07 UT to 20:44:43 UT.  Unlike in \cite{graham2020} and \cite{xu2023}, the components of the double-Gaussian models here are not spectrally resolved and do not evolve significantly during the ${\sim}$2.5 minutes of observations. 

These measures of line width quantify the broadening and asymmetry of the \ion{Ca}{II}~H and H$\epsilon$ lines from the ViSP and highlight that spectral line shape is spatially dependent even during non-optimal observing conditions.  In Section \ref{sec:results} we compare observations of this spectral range to RHD simulations in order to investigate the physical quantities affecting the width and shape of these two lines. 

\section{Modeling}\label{sec:results}

The lines of the \ion{Ca}{II} ion are often studied in the context of solar flare physics and are known to exhibit optically-thick properties similar to \ion{Mg}{II} in the quiet Sun (e.g., \cite{carlsson1997,kerr_2016,kuridze2017,bjorgen2018}).  In the case of \ion{Ca}{II}~H specifically, the nearby H$\epsilon$ line can be used to place additional simultaneous constraints on models.  In this section we present the first comparison of a 1D RHD simulation with the widely-used RADYN and RH codes to solar flare observations of \ion{Ca}{II}~H and H$\epsilon$ from DKIST. For more detailed descriptions of the RADYN code, see \cite{abbett1999,allred2015,carlsson2023,kowalski2024b}.  Select models that are compared directly to ViSP data in this work are listed in Table \ref{tab:modsum}. For simplicity, we will refer to the model IDs shown in the first column of Table \ref{tab:modsum}, terminology which we adopt from \cite{lorincik_2025}.

\begin{table}
\caption{RADYN model details}\label{tab:modsum}
\resizebox{\textwidth}{!}{\begin{tabular}{|c|c|c|c|c|c|c|c|}
\hline
Model ID & $E_{tot}$ [$\mathrm{erg\;cm^{-2}}$] & $F_{max}$ [$\mathrm{erg\;cm^{-2}\;s^{-1}}$] & $E_{c}$ [$\mathrm{keV}$] & $\delta$ & $t_{heat}$ [s] & $t_{relax}$ [s] & HP\\
\hline
EB1 & $1 \times 10^{12}$ & $5 \times 10^{10}$ & 15 & 8 & 20 & 480 & C \\ 
EB2 & $3 \times 10^{11}$ & $3 \times 10^{10}$ & 15 & 3 & 20 & 30 & T  \\ 
EB3 & $1 \times 10^{12}$ & $1 \times 10^{11}$ & 15 & 3 & 20 & 30 & T  \\ 
EB4 & $3 \times 10^{11}$ & $3 \times 10^{10}$ & 15 & 8 & 20 & 30 & T  \\
TC1 & $5 \times 10^{11}$ & N/A & N/A & N/A & 100 & 0 & C  \\ 
\hline
\end{tabular}}
{\raggedright \tiny \textit{Note. Model IDs and parameters corresponding to 5 RADYN+RH models compared to  DKIST spectra in this work.  We refer to models in this work using the labels in the first column.  Model parameters listed include the total energy $E_{tot}$, maximum flux $F_{max}$, low-energy cutoff $E_c$, spectral index $\delta$, heating time $t_{heat}$, relaxation time $t_{relax}$, and heating profile HP, which is either triangular (T) or constant (C).}\par}

\end{table}

\subsection{Beam-heated simulations}
\subsubsection{Estimating beam parameters with the F-CHROMA grid}\label{sec:RADYNdescrip}

The observed hard X-ray spectrum of a flare is normally used to deduce the spectral properties of the electron beam (namely, the flux $F$, low-energy cutoff $E_c$, and spectral index $\delta$) that heats the flare atmosphere in a RADYN simulation.  Unfortunately, the STIX instrument (\cite{krucker2020b}) did not collect observations of the flare ribbons because Solar Orbiter was located almost completely on the opposite side of the Sun\footnote{\url{https://solarorbiter.esac.esa.int/where/}}.  Fermi/GBM spectra also can not in this case be used to constrain beam properties.  This is because there is spurious background emission, not related to the flare, at energies above 25 keV.  This feature is present in all sunward-facing detectors and obscures the flare-only HXR spectrum.  Since no single detector is free from this effect, analysis of GBM spectra to diagnose beam properties is unreliable.

Therefore, we interrogate the F-CHROMA grid \citep{carlsson2023} of electron-beam-heated RADYN models to trial a range of values for $\delta$, $E_c$, and the total beam energy $E_{tot}$ and estimate appropriate simulation beam parameters.  In Section \ref{hepsilon_constrain} we also use models from the F-CHROMA grid to investigate the impact of electron density $n_e$ on the width of H$\epsilon$.  The peak flux densities from the F-CHROMA grid studied here in detail are relatively small.

The F-CHROMA grid of models is widely used to compare flare observations to simulations with a range of electron beam parameters (e.g., \cite{sadykov2020,young2024,monson2024}). The F-CHROMA grid essentially uses the version of RADYN presented in \cite{allred2015} with some differences in atomic levels and in the initial atmosphere.  \cite{carlsson2023} describes the lines modeled in F-CHROMA simulations, including those of a six-level hydrogen atom, the \ion{Ca}{II} ion, and the \ion{He}{I} and \ion{He}{II} ions.  The grid varies the total energy ($E_{tot} = \text{3} \times \text{10}^{\text{10}},\;\text{10}^{\text{11}},\;\text{3}\times \text{10}^{\text{11}},\;\text{10}^{\text{12}}\; \text{erg}\; \text{cm}^{\text{-2}}$), low-energy cutoff ($E_c =$ 10, 15, 20, 25 keV), and power law index ($\delta =$ 3, 4, 5, 6, 7, 8) of the electron beam.  The distribution of pitch angle of the injection electrons is Gaussian with a standard deviation of $\mu = cos\;\theta = \text{0.1}$. This parameter range is {established according to typical values for electron beams in solar flares (e.g., \cite{holman2003,hannah2011,milligan2014,Warmuth2016,carlsson2023}).  For a detailed description of the F-CHROMA starting atmosphere, see \cite{carlsson2023}.  In all F-CHROMA models, we are limited to a half-loop length of 10 Mm.  From the AIA imagery (Figure \ref{sdocontext}(c)), the activated flare loops at the time of the DKIST observations may be larger (up to $\sim\text{20}\;\mathrm{Mm}$).  We discuss this in Section \ref{sec:disc}.

The beam energy flux is injected in a triangular temporal pattern, with a linear 10 s increase to maximum flux at t = 10 s and a linear 10 s decrease to zero beam flux at t = 20 s followed by a 30 s relaxation period. We note that although we often reference the electron beam flux (similarly to e.g. \cite{kowalski2024}, though we refer specifically to the peak beam flux $F_{max}$), we will also refer to the total beam energy $E_{tot}$, which is varied in the F-CHROMA grid.  For the triangular injection of energy in each F-CHROMA model (though not the custom models described below with constant energy injection), the relationship between these two is $E_{tot} = F_{max}\times\text{10}$~s. The relaxation phase is included in the F-CHROMA models to facilitate investigation of the persistent emission in chromospheric spectral lines after beam injection has ceased, a possible scenario for our flare-decay-phase DKIST observations.  

From our investigation of F-CHROMA models, we find that simulations with a relatively soft power law index $\delta=$ 8, $E_c=$ 15 keV, and moderate peak flux from $\text{3}\times\text{10}^{\text{10}}\;\mathrm{erg\;cm^{-2}\;s^{-1}}$ ($E_{tot}=\text{3}\times\text{10}^{\text{10}}\;\text{erg}\;\text{cm}^{\text{-2}}$) to $\text{10}^{\text{11}}\;\mathrm{erg\;cm^{-2}\;s^{-1}}$ ($E_{tot}=\text{10}^{\text{12}}\;\text{erg}\;\text{cm}^{\text{-2}}$) reproduce a single-peaked Ca II H line in the late relaxation flare phase.  From interrogation of the SDO/AIA 1600\AA\ data, we find that the region of the flare ribbon captured by ViSP is heated roughly 8 min before ViSP observations commence at 20:42:07 UT.  Therefore, to compare directly to ViSP observations we generate a new, custom long-duration RADYN model (EB1) with the parameters $\delta=8$, $E_c=$ 15 keV, and a 20~s injection of energy at a constant flux of $F=\text{5}\times\text{10}^{\text{10}}\;\mathrm{erg\;cm^{-2}\;s^{-1}}$, within the range of peak flux values referenced above.  Energy injection in model EB1 is followed by 580s of relaxation.  We choose the generic, constant energy injection rate in this model due to the ambiguity introduced by the lack of reliable HXR observations during the flare.  The loop length in the input atmosphere of this model was 11~Mm.  The simulated spectra analyzed below are taken at $t=500$~s in simulation EB1, 8 min after the cessation of beam energy injection.

\subsubsection{Evolution of a beam-heated RADYN atmosphere}\label{chosenone}

Here we describe the relevant details in the atmospheric evolution of the long-duration, beam-heated model EB1. In Figure \ref{model_evolve} we show the evolution of the temperature, electron density, macroscopic velocity, and the \ion{Ca}{II}~H line profiles at select times in this simulation. Panels (a), (b), and (c) show the evolution of the loop temperature $T$, field-aligned velocity $v_z$, and electron density $n_e$ at several stages of the simulation.  Although the coronal temperature at loop-top begins at 3 MK, after beam energy injection this temperature increases to over 10 MK.  Energy injection produces a dense chromospheric condensation with $n_e>\text{10}^{\text{14}}\;\text{cm}^{\text{-3}}$.  Due to the impact of the electron beam, the transition region moves downward from ${\sim}$ 1.2 Mm to ${\sim}$ 0.75 Mm.  During and directly after the beam energy injection, \ion{Ca}{II}~H exhibits a strong redward asymmetry which then evolves toward an un-shifted stationary component (panels (d), (e)-(l)).  During the relaxation period starting at t = 20 s and until the end of the simulation at t = 600 s, the transition region moves from ${\sim}\text{0.75}\;\mathrm{Mm}$ to $>\text{1}\;\mathrm{Mm}$ and evolves more slowly, with peak upper chromospheric electron densities decreasing from $n_e\approx\text{10}^{\text{13}}\;\text{cm}^{\text{-3}}$ to $n_e\approx\text{2}\times\text{10}^{\text{12}}\;\text{cm}^{\text{-3}}$ at t = 500 s.  Early in the simulation there is a lower, less compact layer in the chromosphere with electron density $n_e\approx\text{10}^{\text{13}}\;\text{cm}^{\text{-3}}$ decreasing to $\text{10}^{\text{12}}\ \text{cm}^{\text{-3}}$, still far in excess of non-flare chromospheric densities.  Ca II H is single peaked until about t = 300 s, when the line becomes clearly double-peaked.

\begin{sidewaysfigure}
\includegraphics[width=1\linewidth]{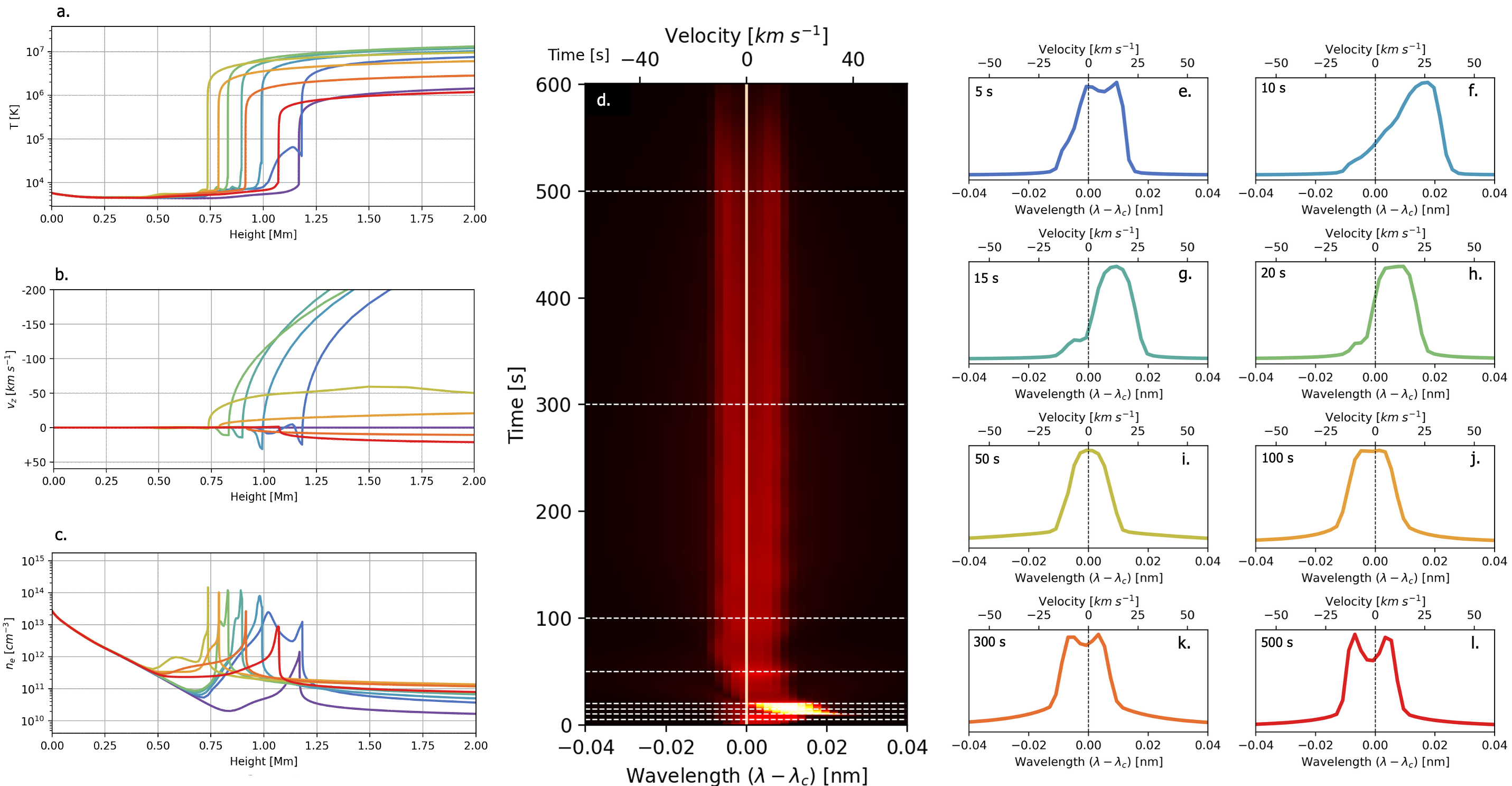}
\centering
\caption{\tiny Evolution of model EB1 prior to application of the RH code.  In order to highlight the variations in the chromosphere and transition region, we show the temperature and electron density profiles up to 2 Mm only. (a) Evolution of temperature profiles throughout the atmosphere at selected times.  The pre-flare atmosphere at $t=0$~s is shown in violet, and progressively warmer colors indicate later simulation phases.  (b) The same as (a), but showing the evolution of vertical plasma velocity. Positive and negative values refer to downflows and upflows, respectively.  (c) The same as (a), but showing the evolution of the electron density.  (d) Spectral-temporal map of \ion{Ca}{II}~H line profile intensity in model EB1.  During the relaxation phase the line exhibits a symmetric, single-peaked emission line profile of lower intensity, eventually turning double-peaked. Horizontal white dotted lines correspond to the time steps indicated in panels (a)-(c) and (e)-(l).  (e)-(l) Ca II H line profiles at times corresponding to the colors in panels (a)-(c). Intensities are normalized for each time step to facilitate better investigation of relative line width.}
\label{model_evolve}
\end{sidewaysfigure}

\subsection{Simulation driven by thermal conduction}
While it is traditionally thought that beam heating is limited to the impulsive phase of the flare, the presence of the effects of beam heating in the decay phase has been posited before (e.g., \cite{qiu_2004}). More recently, \cite{krucker2020} and \cite{chen2024} observed non-thermal radio emission, indicative of particle acceleration and trapping, in the gradual decay phase of multiple events. However, in the absence of hard X-ray signatures, conduction from the coronal reconnection site is another possible mechanism for flare heating, particularly after the flare impulsive phase.  As a test, we also compare the DKIST data to a simple RADYN model heated only by thermal conduction (model TC1).  This model runs for 100~s and includes constant heating at $\text{5}\;\text{erg}\;\text{cm}^{\text{-3}}\;\text{s}^{\text{-1}}$ from $h=$ 1.4 Mm to $h=$ 11 Mm for 100~s.\footnote{The duration of heating and location of in situ energy deposition can not be reliably constrained with observations.  Regarding the timing of energy injection, we choose 100 s in order to balance reasonable values for the instantaneous in situ flux with the total energy delivered to the atmosphere. Regarding the spatial distribution of energy injection, the atmospheric evolution is relatively ambivalent to the spatial distribution of heating due to the redistribution of energy in a conduction-driven scenario (e.g., \cite{polito_2018}), so long as it is not imposed at the footpoints.  The homogeneous in situ heating from 1.4 Mm to 11 Mm is therefore a reasonable choice.}   This delivers $E_{tot}\approx\text{5}\times\text{10}^{\text{11}}\;\text{erg}\;\text{cm}^{\text{-2}}$ to the atmosphere over the course of the entire simulation.  This is half the total energy delivered in model EB1, but is reasonable for the entire observed emission of this C-class flare and in the middle of the range of derived values for $E_{tot}$ from investigation of the F-CHROMA grid.

In the following we compare model TC1 at t = 90 s to the DKIST observations, at which point $\text{4.5}\;\text{erg}\;\text{cm}^{\text{-2}}$ of energy has been delivered to the simulated atmosphere. From analysis of the model, we find that this choice is not crucial to the behavior of Ca II H as long as the atmosphere has sufficiently redistributed energy following the initial injection of conductive flux.  During the latter stages of the simulation, Ca II H is consistently symmetric and single peaked with only gradual changes in absolute intensity.  In the future, the spatial and temporal details of heating in this simple model could be varied to study the implications in resultant spectra.

\subsection{The RADYN+RH bridge}\label{sec:RHdescrip}
Notably, the RADYN models are produced using a 6-level hydrogen atom, which is insufficient for synthesis of the H$\epsilon$ $n_j=$ 7 to $n_i=$ 2 transition.  Incorporation of a 7-level hydrogen atom into RADYN is not preferred due to RADYN's handling of overlapping transitions such as \ion{Ca}{II}~H and H$\epsilon$.  Therefore, we feed snapshots of the RADYN atmosphere details, non-equilibrium electron densities, and hydrogen population densities from F-CHROMA to the statistical equilibrium 1D RH code, which can include a 20-level hydrogen atom and handles overlapping transitions self-consistently \citep{uitenbroek_2001,uitenbroek2002}.  Following the findings of e.g. \cite{zhu2019}, \cite{yadav_2021}, \cite{sainzdalda_2023}, and \cite{kerr_2024}, for some models we adjust the \textit{ad hoc} microturbulence parameter $v_{turb}$ in the input RH atmosphere file to $\text{7}\;\text{km}\;\text{s}^{\text{-1}}$ from the default $\text{2}\;\text{km}\;\text{s}^{\text{-1}}$.  We comment on this choice of $v_{turb}$ further in Section \ref{sec:disc}. Hereafter we refer to the resulting ``bridge'' simulation as RADYN+RH. 

The hydrogen broadening prescription calculated using the modifications of \cite{tremblay2009} and occupational probability formalism of \cite{hummer1988}  (TB09+HM88) was incorporated into the RH code by \cite{kowalski2017b}.  RH includes partial frequency redistribution (PRD) effects.  Although the RH code is in statistical equilibrium (SE), this assumption does not significantly affect the line profile of higher-order hydrogen line transitions when non-equilibrium (NEQ) electron densities from RADYN snapshots are used as input to RH.  A direct comparison of RADYN and RH calculations of H$\epsilon$ is not possible.  However, Figure 9 of \cite{kowalski2022a} shows a comparison of the RADYN and RH calculations of H$\gamma$, which suggests that the SE assumption is likely sufficient for higher-order Balmer series lines.  In Appendix C we demonstrate that the SE assumption also does not substantially affect the line width of Ca II H during the simulation relaxation phase.

\subsection{Comparing RADYN+RH simulations to DKIST observations}\label{sec:mod_comp}

Here we analyze the resultant \ion{Ca}{II}~H and H$\epsilon$ RADYN+RH model spectra. We focus only on the ViSP spectra taken at 20:42:07 UT. For appropriate comparison to ViSP spectra, we convolve all RADYN+RH spectra with a Gaussian point spread function (PSF) of width 5 pm, empirically determined during intensity calibration (Appendix \ref{sec:app1}) to be an appropriate instrument PSF for the ViSP. Because Ca II H and H$\epsilon$ may not be optically thin under flare conditions, their emission is likely coupled to the pre-flare atmosphere.  We subtract the non-flare spectrum from both observations and simulations to isolate the emission line signature due to the flare.  This removes distracting discrepancies between the simulated and observed pre-flare atmosphere.  The salient features of spectral line shape investigated here (line width, relative line intensity) are present with or without subtraction of the non-flare spectrum.  For example, the Ca II H line profile at ribbon center before subtraction (Figure \ref{kernzoom}(a)) and after subtraction (Figure \ref{kernzoom}(c1), green line) are morphologically indistinguishable.

The RADYN+RH bridge synthesizes emergent spectral intensity at several viewing angles.  In our analysis we choose the simulation with $\mu_{sim} =$ 0.5, close to $\mu_{obs} =$ 0.48.

The results for models EB1 and TC1 are shown in Figure \ref{fig:fullsimtest}(a), and for all models in Table \ref{tab:modsum} in Figure \ref{fig:fullsimtest}(b). Here we compare profiles normalized to the intensity of H$\epsilon$ in order to focus on a comparison of the modeled and observed line width.\footnote{The peak spectral line intensity is another potential modeling constraint, but there are many factors (principally the spatial resolution of observations compared to a simulated flare kernel in a 1D model, but also the presence of a central reversal, the decomposition of a spectral line into distinct components, or the precise timing of observations) that obfuscate the comparison of observed and modeled peak line intensity.  \cite{osborne_2022} investigate the impact of spatial resolution on line intensity in a realistic, 3D scenario, which they suggest must be considered when comparing observed to simulated line profiles.}  Although the output using the default RH input atmosphere (with $v_{turb}=\text{2}\;\text{km}\;\text{s}^{-1}$) is too narrow; blue and red dashed lines, Figure \ref{fig:fullsimtest}(a), models EB1 and TC1 produce broad, relatively symmetric Ca II H profiles with a blue wing that reflects the observed line width after adjustment of $v_{turb}$.  The resultant line profile from EB1 is double-peaked.  Both models significantly underestimate the width of Ca II H in the red wing.  Apparent overestimation of the blue wing line width by model TC1 is mostly a result of normalization to the intensity of H$\epsilon$.

The relative intensity of Ca II H to H$\epsilon$ is overestimated by model TC1 and underestimated by model EB1.  However, both models reproduce the width of H$\epsilon$ more successfully (with only a slight underestimation in the blue wing in some models), as we discuss further in Section \ref{hepsilon_constrain}. 
\begin{figure}
\includegraphics[width=1\linewidth]{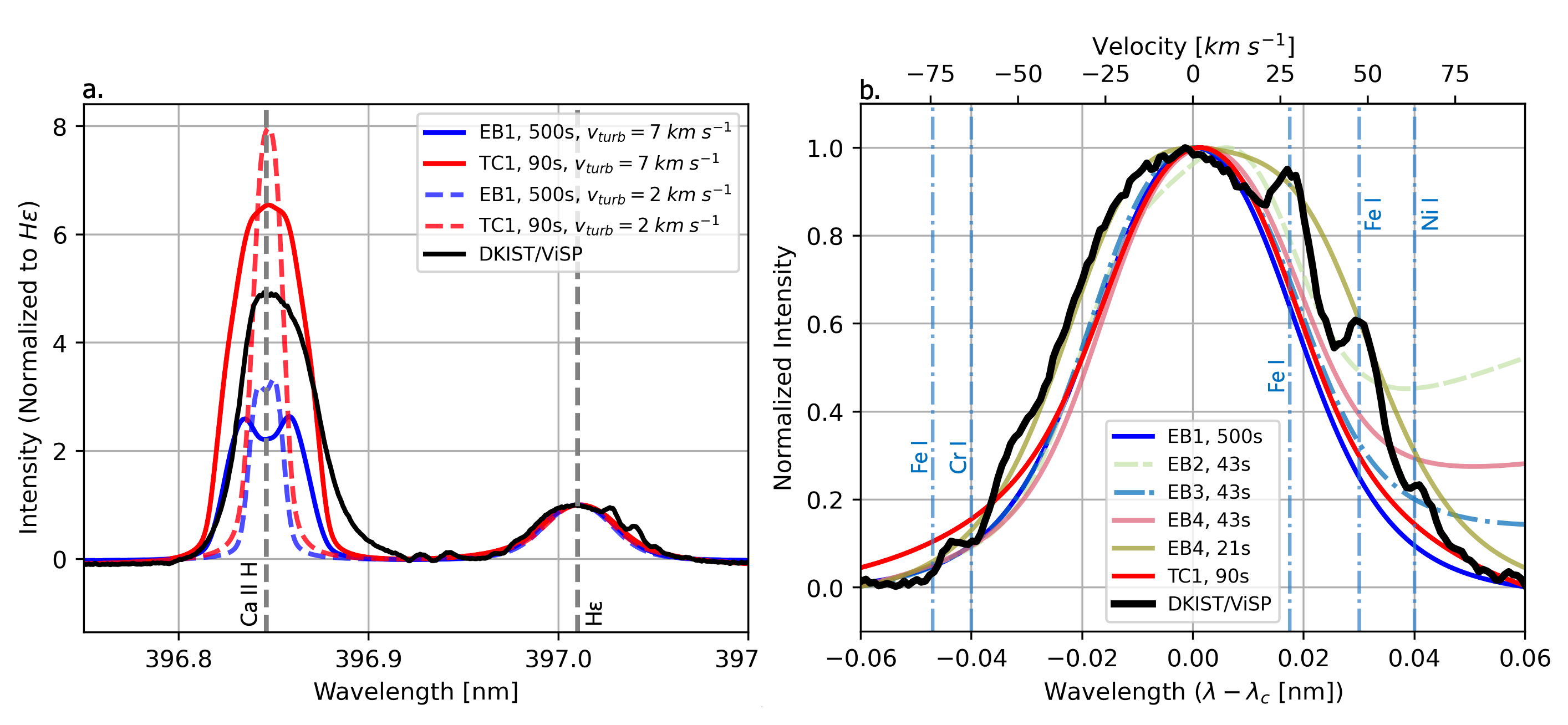}
\centering
\caption{\tiny (a) A comparison of RADYN+RH simulated \ion{Ca}{II}~H and H$\epsilon$ lines to observations.  (b) A comparison of ViSP observations to the modeled H$\epsilon$ profiles that are notably in emission in panel (a), using $v_{turb}=2\;\mathrm{km\;s^{-1}}$ in the input RH atmosphere file.  We note the locations of several other lines within the spectral range of H$\epsilon$ from \ion{Fe}{I}, \ion{Fe}{I}, and \ion{Ni}{I}.  The two \ion{Fe}{I} lines in the red wing are deeper in the quiet Sun than the flare spectra, giving the impression of an emission line when the pre-flare is subtracted.  Intensity values include pre-flare subtraction and are normalized to the maximum intensity of H$\epsilon$ in order to easily compare the widths of observed and modeled lines. The observed line profiles from ViSP at ribbon center at 20:42:07 UT are shown in black.}
\label{fig:fullsimtest}
\end{figure}

\subsection{The effect of \texorpdfstring{$n_e$}{electron density} on H\texorpdfstring{$\epsilon$}{e} line width}\label{hepsilon_constrain}

In this section we perform an experiment using all models listed in Table \ref{tab:modsum} to demonstrate the range of values for $n_e$ that can produce similar widths in H$\epsilon$ (Figure \ref{fig:fullsimtest}(b)).  We study models EB1 at 500 s, EB2, EB3, and EB4 at t = 43 s, EB4 at t = 21 s, and model TC1 at t = 90 s.  We include models EB2 and EB3, with harder spectral index $\delta=$ 3, to widen the grid of atmospheric conditions studied; this is not intended to reflect the real atmosphere producing the observed H$\epsilon$. Note that the atmospheres used to produce models EB2-4 are taken directly from the F-CHROMA grid, and therefore are only allowed to evolve for 30~s after beam injection; these are included in order to explore a wider range of model atmospheres and their effect on H$\epsilon$ line width.   Since the physical significance of $v_{turb}$ is unclear (e.g., \cite{kowalski2024}), we use the default $v_{turb}=2\;\text{km}\;\text{s}^{\text{-1}}$ for the comparison in Figure \ref{fig:fullsimtest}(b). This choice has no significant effect on the results of this analysis since $v_{turb}$ variations do not significantly alter the modeled width of H$\epsilon$ (as shown in Figure \ref{fig:fullsimtest}(a)).

We note that the purpose of this investigation is to demonstrate the impact of electron density on the width of H$\epsilon$ for several simulations, readily available with the F-CHROMA grid.  The 50~s duration of the F-CHROMA models means that a direct time comparison to the late-decay-phase ViSP spectra is not possible.  We choose t = 43 s for models EB2-4 because (i) all three calculations with RH converge at this time step, which is not the case later in the simulation, and (ii) this is the last time step for which Ca II H is single-peaked in all three RADYN simulations prior to evolution with RH.  Although we do not compare the resultant Ca II H profiles in this section, we can use this as a constraint to choose an appropriate time step in the absence of the H$\epsilon$ line, which is not calculated by RADYN.  EB4 at t = 21 s (directly after energy injection ceases) is included to investigate differences in H$\epsilon$ line width under different conditions.

As indicated in Figure \ref{fig:fullsimtest}(b), in the red wing of H$\epsilon$ there are two \ion{Fe}{I} absorption lines that appear slightly more intense at flare-time compared to the non-flare spectrum, giving the impression of emission lines superimposed on the H$\epsilon$ spectrum. We identify these lines using the Moore solar spectrum \citep{moore1965} and are uncertain of the precise origin of these lines in the flaring solar atmosphere.  Taking the \ion{Fe}{I} lines into consideration, the observed width of H$\epsilon$ in the red wing is slightly less than it appears.  Therefore, model EB1 at 500~s, models EB2--4 at 43~s, and model TC1 at t = 90 s only slightly underestimate the width of H$\epsilon$ in both the red and the blue.  EB4 at t = 21 s captures the width in the blue wing well, but after considering the enhanced intensity from the \ion{Fe}{I} emission lines, it slightly overestimates the observed width in the red wing. Note that the continuum in the red and blue wings of H$\epsilon$ in some models is not at the same intensity; this is due to the width and intensity of Ca II H in the corresponding model.  In these models we focus only on the observed width in the near red-wing of H$\epsilon$, since investigation of the far wing is obfuscated.

We derive $n_e$ at the location of the peak contribution to the emergent intensity of H$\epsilon$ from the output RADYN+RH atmosphere. In all models, the maximum contribution to H$\epsilon$ occurs in the upper chromosphere, where $n_e$ experiences a sharp peak over a relatively small range of heights.  We list the corresponding electron densities in Table \ref{tab:elecdens}.  During the late relaxation phase, the simulations for which H$\epsilon$ is in emission demonstrate a range of electron densities spanning $n_e\approx\text{2}\times\text{10}^{\text{12}}\;\text{cm}^{\text{-3}}$ to $8\times\text{10}^{\text{13}}\;\text{cm}^{\text{-3}}$.  As discussed below, some models (notably EB1 at 500~s and EB4 at 21~s) have prominent contributions to the width of H$\epsilon$ from lower layers in the chromosphere, suggesting that upper chromospheric $n_e$ is not the sole predictor for the width of H$\epsilon$.  In model EB4 at 21~s, for example, there is a prominent stationary layer at slightly lower density that contributes significantly to the formation of H$\epsilon$.  We discuss these patterns and the complications of H$\epsilon$ line formation in Section \ref{sec:disc} below.

\begin{table}
%\centering
\begin{threeparttable}

\caption{Condensation electron densities}
\label{tab:elecdens}
\begin{tabular}{|c|c|c|c|}
\hline
Model ID & Time $[\mathrm{s}]$& $n_e$ [$\mathrm{cm^{-3}}$] & Figure\\ \hline
EB1  & 500 & $9\times10^{12}$ & \ref{fig:fullsimtest}(a),(b)\\
EB2  & 43 & $2\times10^{12}$ & \ref{fig:fullsimtest}(b)\\
EB3 & 43 & $4\times10^{13}$ & \ref{fig:fullsimtest}(b)\\
EB4  & 43 & $3\times10^{13}$ & \ref{fig:fullsimtest}(b)\\
EB4  & 21 & $2\times10^{13}$ & \ref{fig:fullsimtest}(b)\\
TC1  & 90 & $8\times10^{13}$ & \ref{fig:fullsimtest}(a),(b)\\
\hline
\end{tabular}\par
\begin{tablenotes}
\item[1] \tiny \textit{Electron densities corresponding to the RADYN+RH models listed in Table \ref{tab:modsum}, with model IDs and references to the corresponding figure.}
\end{tablenotes}
\end{threeparttable}
\end{table}

In summary, here we have investigated the modeled line profiles of \ion{Ca}{II}~H and H$\epsilon$ from RADYN+RH simulations.  We demonstrated that while the line width of H$\epsilon$ is fairly well constrained, the simulated Ca II H line width is significantly underestimated. In the following section we discuss the implications of these findings.

\section{Discussion}\label{sec:disc}
\subsection{Interpretation of line asymmetry measurements}

At the ribbon center, the DKIST spectra show asymmetric, single-peaked \ion{Ca}{II}~H emission line profiles with a wider red than blue wing (Figure \ref{kernzoom}).  For optically thin chromospheric lines like \ion{Si}{IV}, asymmetries are often explained using a chromospheric evaporation/condensation model, where red- and blue-shifted lines or line components represent down- and up-flowing chromospheric material respectively (e.g., \cite{ashfield2022,xu2023}).  This method has also been applied to optically thick lines such as the \ion{Mg}{II} triplet when these lines exhibit clear Gaussian components  (e.g., \cite{graham_cauzzi2015,graham2020}).  

We characterize the line profile shape of \ion{Ca}{II}~H using a similar method in Section \ref{sec:CaIIvary} (Figure \ref{kernzoom}(c4-c5)) and find that the single-peaked \ion{Ca}{II}~H line is well-represented by a double-Gaussian model with a blue-shifted component at ${\sim}$ --16 to --19 $\text{km}\;\text{s}^{\text{-1}}$ and a red-shifted component at ${\sim}$ 6 to 11 $\text{km}\;\text{s}^{\text{-1}}$.  The velocity shift of these components varies across the flare ribbon, with a slightly larger redshift at the ribbon trailing edge during the flare gradual phase (see the red profile in Figure \ref{kernzoom}(c1)), possibly indicative of coronal rain (e.g., \cite{lacatus2017,reep2020}), though we do not test this possibility here.  We find that at the ribbon center, the blue component of \ion{Ca}{II}~H slightly increases in velocity and the red component decreases in velocity over the duration of observations.  

We caution that evolution of the components of the line profile should not necessarily be interpreted as evidence of mass motion.  First, the data are limited to seven scans covering only ${\sim}$2.5 minutes of the gradually-evolving decay phase (Figure \ref{fig:goes_summary}) and are subject to variable seeing.  Also, the physical interpretation of a double-Gaussian model is less certain for observations of optically thick lines like \ion{Ca}{II}~H without spectrally distinct Gaussian components.  Therefore, despite the accuracy of the simple double-Gaussian model in decomposing the observed emission line, in this paper we hesitate to claim an evaporation-condensation model, indicated by double-Gaussian fitting, as an explanation for the line profile width.   

\subsection{Remaining model-data discrepancies in the \ion{Ca}{II}~H emission line}

The width of Ca II H in the red wing is significantly underestimated by the RHD models investigated here (Figure \ref{fig:fullsimtest}(a)).  \cite{zhu2019} found a qualitatively similar (though symmetric) discrepancy in their analysis of the \ion{Mg}{II} h, k, and subordinate lines.  They found that a multiplicative factor of 30 on the quadratic Stark-Lo Surdo line width was required to resolve this difference, suggesting that the RADYN and RH codes neglect some relevant Lorentzian effect on the line width.

As also demonstrated here, non-Lorentzian sources of broadening are also important when accurately resolving the line width.  For example, the ad hoc microturbulence parameter has a Gaussian effect on the line width.  \cite{kerr_2024b} found that microturbulent velocities of $\text{7.5--10}\;\text{km}\;\text{s}^{\text{-1}}$ are appropriate in a flare atmosphere. \cite{zhu2019} investigated the effect of increasing microturbulence manually at different heights (up to $\text{50}\;\text{km}\;\text{s}^{\text{-1}}$ at lower heights) and found that large values of microturbulence produced an unrealistically enhanced line core in \ion{Mg}{II} lines and did not resolve the discrepancy between simulations and observations in the far wings of the line.  Modeled profiles with a uniform microturbulence of 5 $\text{km}\;\text{s}^{\text{-1}}$  in their work seem to better reproduce the \ion{Mg}{II} line core, although they do not discuss this extensively.  Using the spectral inversion code STiC \citep{delacruzrodriguez_2019} to quantify microturbulence among other atmospheric parameters, \cite{sainzdalda_2023} found an atmospheric microturbulence of $\xi = \text{5--15}\;\text{km}\;\text{s}^{\text{-1}}$.  \cite{yadav_2021} inverted ground-based \ion{Ca}{II} 8542 \AA\ and \ion{Ca}{II} K flare observations, finding values of microturbulence ranging from a few $\text{km}\;\text{s}^{\text{-1}}$ to $\text{5--10}\;\text{km}\;\text{s}^{\text{-1}}$ in the upper chromosphere.  Our choice of $v_{turb}=\text{7}\;\text{km}\;\text{s}^{\text{-1}}$ generally explains the discrepancy in predicted blue wing width for Ca II H, but fails to resolve the red wing discrepancy. This remaining issue may be explained by spatial resolution effects, by constraints placed on Lorentzian effects such as in \cite{zhu2019}, or by some other physical effect not included in the model. We note also that given the clear spectral variability across the flare ribbon in data with suboptimal observing, it is likely that resolution effects may also contribute to the unexplained elevated Ca II H red wing.

Finally, we note that the relative intensity of Ca II H to H$\epsilon$ is underestimated by the electron beam-heated model EB1, and overestimated by the conduction-driven model TC1 (Figure \ref{fig:fullsimtest}(a)).  This could be due to uncertainty in the beam parameters or input conductive flux.  Alternatively, the discrepancy in relative line intensity may suggest that a combination of both heating mechanisms is more appropriate for comparison to decay-phase flare spectra.  A more comprehensive modeling investigation considering these points is necessary to address this issue, but outside the scope of this work.  Such a study may include a comparison of DKIST flare data to RADYN simulations heated by other mechanisms (e.g., proton beams, Alfv\'enic waves) and constraints placed by direct observations of the HXR spectra, which are unreliable for this flare.

\subsection{Chromospheric electron density and the formation height of H$\epsilon$}

The presence of H$\epsilon$ in the same spectral window as \ion{Ca}{II}~H is a novel feature in modern high-resolution solar flare spectra, although the lines were observed often during flares in the mid-to-late-$\text{20}^{th}$ century (e.g., \cite{Svestka1976_rcc}).  The proximity of \ion{Ca}{II}~H to H$\epsilon$ presents the opportunity to produce a self-consistent RHD simulation that simultaneously captures the dynamics of \ion{Ca}{II}~H and H$\epsilon$ lines, subject to distinct constraints in atmospheric and beam parameters. In high-energy flares observed on both the Sun and other stars, H$\epsilon$ is often very broad, with a blue wing blending heavily with the entire Ca II H line.  The relatively narrow decay-phase H$\epsilon$ line profile in the flare reported here is therefore notable, as is the consistency of the models employed in reproducing the narrow H$\epsilon$ line width (Figure \ref{fig:fullsimtest}(b)). We determine that the chromospheric layers in the late relaxation phase contributing to broadened emission in H$\epsilon$ have electron densities in agreement with previous work (e.g., \cite{neidig1984,johnskrull1997}, and many others).  These densities are a factor of ${\sim}$ 10 lower than the electron densities that lead to broadening in the low-$n_j$ Balmer lines in \cite{kowalski2022a}, though the predictions in that work are intended for impulsive phase kernels, when larger beam flux densities are inferred.

Since the reproduced line widths in these models are all fairly consistent despite a wide range of electron densities, factors other than the upper chromospheric density seem to be relevant to the broadening of H$\epsilon$. For example, as shown in Figure \ref{fig:fullsimtest}(b), model EB4 at t = 21 s (one second after the cessation of beam energy injection) overestimates the width of H$\epsilon$ compared to the other models shown despite a ${\sim}$30\% lower electron density at t = 21 s compared to t = 43 s.  At t = 21 s in model EB4, there is a significant contribution to the line width (particularly in the wings) from a layer lower in the chromosphere with $n_e\approx\text{9}\times\text{10}^{\text{12}}\;\text{cm}^{\text{-3}}$.  Further, model EB1 at 500~s and EB2 at 43~s produce H$\epsilon$ lines with approximately the same width despite very different electron densities of $n_e\approx\text{9}\times\text{10}^{\text{12}}\;\text{cm}^{\text{-3}}$ and $n_e\approx\text{2}\times\text{10}^{\text{12}}\;\text{cm}^{\text{-3}}$ respectively - the latter is more than a factor of 10 lower than the $n_e$ in model EB4 at 21~s.  Model TC1 at t = 90 s has an upper chromosphere electron density $n_e=\text{8}\times\text{10}^{\text{13}}\;\text{cm}^{\text{-3}}$, much higher than the other models, but produces a H$\epsilon$ line width very similar to that of model EB1.

We therefore suggest that in addition to upper chromospheric layers, contributions from lower layers and the height of the upper chromosphere may also play a significant role in the formation of H$\epsilon$. Predictions of Balmer series line broadening from \cite{kowalski2022a} indicate that, while low-$n_j$ Balmer lines form mostly in upper chromospheric layers with $n_e>\text{10}^{\text{14}}\;\text{cm}^{\text{-3}}$ during a flare owing to their large optical depths, the less-optically-thick high-$n_j$ lines (specifically, $n_j=\text{12--16}$) may form largely in lower, stationary layers with lower $n_e$ and small curve-of-growth broadening enhancements.  Although it has a lower $n_j$ than the lines investigated in that work,  H$\epsilon$ ($n_j =$ 7) may also form partially in lower atmospheric layers.  Further investigation into the factors involved in producing an H$\epsilon$ line during the relaxation phase of a flare simulation is required in order to explain why upper chromosphere electron density alone does not seem to predict the width of H$\epsilon$. 

\subsection{Additional avenues for further investigation}

Future work that may resolve the discrepancy between model and data includes investigation of a larger grid of RHD models. As discussed at length in e.g. Section 3 of \cite{kerr2023r}, Alfv\'enic waves (e.g., \cite{kerr_2016,reep_2016}), conductive heat fluxes (e.g., \cite{polito_2018,allred_2022}), and even high flux non-thermal electron beams (e.g., \cite{kowalski2017a}) or proton beams (e.g., \cite{prochazka_2018,sadykov2023,kerr_2023_oz,kerr2026}) all produce simulated spectral lines consistent with certain aspects of observed flare spectra.  A combination of these heating effects could be responsible for the remaining observed discrepancies in synthesizing \ion{Ca}{II}~H, particularly in the gradual decay phase.  The presence of H$\epsilon$, which broadens due to different mechanisms than \ion{Ca}{II}~H, in this spectral window may help provide simultaneous additional constraints when testing the effects of these heating sources. 

Additionally, the F-CHROMA models in particular are limited in their assumptions regarding the starting atmospheric structure.  For example, although the 10 Mm loop length used in the F-CHROMA grid (and 11 Mm loop length in models EB1 and TC1) appears close to the loop length estimated from AIA images, this parameter should be varied in future work to explore the effects on the beam heating and line evolution and potentially resolve remaining model-data discrepancies. Thermal conduction may be more important in longer loops due to the increased beam-stopping power of the coronal densities (e.g., \cite{polito_2018,kerr2022r}).

In the future we will also perform a similar comparison of RHD simulations to impulsive phase flare observations to further address remaining questions in chromospheric line broadening observed in both solar and stellar flare studies (e.g., \cite{kowalski2022a,notsu2024,notsu2025}, among many others).  Model \ion{Ca}{II}~H and H$\epsilon$ lines during or immediately following the phase of beam energy injection are much more intense than observed in the decay phase here and are typically double-peaked or redshifted, much different from the persistent single-peaked, slightly asymmetric observed line (see Figure \ref{fig:fullsimtest}(a) for an example of the double-peaked profile).  Impulsive-phase observations would clearly determine whether the significant asymmetries predicted by the model during the energy injection phase are accurate. Additionally, the availability of co-temporal hard X-ray observations from SolO/STIX, Fermi/GBM, or the Hard X-ray Imager on the Advanced Space-based Solar Observatory is crucial for accurate assessment of the injected electron beam parameters.

We note that for optically thick lines, the pre-flare atmosphere has significant bearing on the flare-time spectral behavior.  Here we have avoided the influence of the pre-flare atmosphere on flare spectra by subtracting the non-flare spectrum from both observations and models.  While this successfully isolates the flare emission in the two lines studied, there are caveats associated with this approach.  For example, the elevated Fe I emission in the red wing of H$\epsilon$ (e.g., Figure \ref{kernzoom}(b1)) is due to this subtraction.  Ideally, RADYN models should be run using a pre-flare atmosphere specifically generated for the flare studied, but the generation of a new starting atmosphere is non-trivial and outside the scope of this work. 

We have yet to investigate the possibility of spatial differences in the formation of these lines relative to the flare ribbon - for example, the core-halo scenarios discussed in Section \ref{sec:intro}.  Additionally, the present work mainly focuses on only one scan of the ViSP.  This is primarily because of the relatively low ViSP map cadence during these observations and also because spectra evolve less rapidly in the decay phase than in the impulsive phase.  Nevertheless, a comprehensive model should capture the time evolution of spectra during all flare phases.  In a future study, we will (i) investigate comprehensive impulsive phase and gradual decay phase observations of \ion{Ca}{II}~H and H$\epsilon$ from DKIST to explicate precise differences in line evolution of an elementary flare kernel in both time and space;\footnote{We note that under better observing conditions, and operating with the short exposures of the intensity-only mode, the spatial resolution of the ViSP may improve while also achieving much higher map cadences.} (ii) collect observations of other Balmer series lines such as H$\alpha$, H$\beta$, or H$\gamma$ simultaneously with \ion{Ca}{II}~H/H$\epsilon$ as suggested by \cite{kowalski2022a}, and (iii) perform a similar comparison to RHD simulations, but with a larger parameter space perhaps also varying the heating mechanisms employed.  Relatedly, the observed red wing at ribbon center may be broadened artificially by a failure to fully resolve the ribbon-center spectrum from the ribbon-trailing-edge spectrum, which shows a more pronounced red wing (Figure \ref{kernzoom}(c1)). This issue may also be addressed by higher-resolution observations (well within the capability of the DKIST).

\section{Conclusions}\label{sec:conc}

In this study, we present the first flare-time observations of \ion{Ca}{II}~H and H$\epsilon$ emission line spectra with the DKIST/ViSP instrument during the decay phase of a GOES X-ray class C6.7 flare at 20:42 UT on 19 August 2022.  We analyze the evolution of the unique spectral window that includes both lines.  With the advent of DKIST observations, these lines will likely be studied frequently in the context of flare physics in the coming years.

 \begin{enumerate}  
     \item We describe the evolution of the \ion{Ca}{II}~H and H$\epsilon$ emission lines during ViSP observations.  We characterize the asymmetry and width of the observed \ion{Ca}{II}~H line. At the flare ribbon center, \ion{Ca}{II}~H is single-peaked with a broad red wing, while H$\epsilon$ is less intense and more symmetric.
     \item We demonstrate that a 1D RADYN+RH bridge simulation can produce emission lines with key characteristics of the observed spectra in the late relaxation phase (t = 500 s), such as the width of H$\epsilon$ and a generally single-peaked Ca II H.  In the absence of reliable hard X-ray spectra for this flare, we infer reasonable beam parameters by comparing spectra from the F-CHROMA database of RADYN models to the ViSP spectra.  The electron-beam model EB1 shows a slight central reversal, as do many of the off-kernel-center Ca II H spectral profiles in Figures \ref{fig:flare_summary_19_august} and \ref{kernzoom}.  The conduction-driven model TC1 displays similar features, although Ca II H is much more intense in that model than in EB1.  In model TC1, Ca II H does not exhibit a central reversal.
     \item All RHD models studied underestimate the width in the red wing of Ca II H significantly, suggesting that improvements to the models and better observing conditions are required to accurately model this line.  As addressed by e.g. \cite{zhu2019}, this is a persistent problem in modeling of non-hydrogenic chromospheric spectral lines in flare codes.  Adjustment of the microturbulence parameter $v_{turb}$ accounts for most of the difference in the blue wing of Ca II H.
     \item The line width of H$\epsilon$ is slightly underestimated by the models in the late relaxation phase and overestimated in the early relaxation phase. The simulations produce a wide range of upper-chromospheric electron densities from $n_e \approx \text{2}\times\text{10}^{\text{12}}\;\text{cm}^{\text{-3}}$ to $\text{8}\times \text{10}^{\text{13}}\;\text{cm}^{\text{-3}}$ despite relatively similar H$\epsilon$ widths across models. Other atmospheric characteristics such as the depth of the upper chromosphere and contributions to line formation from lower layers seem to have a strong effect on H$\epsilon$ line width.  The flare-time formation of H$\epsilon$ may differ from that of lower-$n_j$ lines like H$\alpha$.  Although it was assumed that H$\epsilon$ would more directly probe the upper chromosphere electron density, the narrow profiles observed here (compared to e.g. the broad H$\epsilon$ reported by \cite{neidig1984} and \cite{Svestka1976_rcc}) suggest otherwise. Peak chromospheric electron density is not a direct predictor of H$\epsilon$ line width.  Further observations of H$\epsilon$ and also lower-order Balmer series lines are required to investigate this possibility.
     
 \end{enumerate}  

The prospect of new observations with DKIST and other instruments, alongside updated forward modeling using improved constraints and state-of-the-art spectral inversion codes, represents a massive opportunity and body of future work for the flare community.  The employment of these techniques in studies of flare data has led to significant advances in our understanding of flare physics.  Our study provides one example of the new science to come and the remaining questions we have left to answer as the community begins to process data from DKIST with new and improved analytical tools. Investigation of impulsive phase flare observations in comparison to RHD simulations during the energy injection phase, coupled with the investigation of heating sources other than an electron beam, is left to future work and may help to resolve the remaining differences between models and observations.

\begin{acknowledgments}
    The research reported herein is based in part on data collected with the Daniel K. Inouye Solar Telescope (DKIST) a facility of the National Science Foundation (NSF).  DKIST is operated by the National Solar Observatory (NSO) under a cooperative agreement with the Association of Universities for Research in Astronomy (AURA), Inc.  DKIST is located on land of spiritual and cultural significance to Native Hawaiian people. The use of this important site to further scientific knowledge is done with appreciation and respect.  Support for this work is provided by the NSF through the DKIST Ambassadors program, administered by NSO and AURA, Inc. Y.N.\ acknowledges funding from  NASA ADAP 80NSSC21K0632, NASA TESS Cycle 6 80NSSC24K0493, NASA NICER Cycle 6 80NSSC24K1194, HST GO 17464, and NSF AGS 1916511. The research leading to these results has received funding from the European Community’s Seventh Framework Programme (FP7/2007-2013) under grant agreement no. 606862 (F-CHROMA), and from the Research Council of Norway through the Programme for Supercomputing.
\end{acknowledgments}

\appendix
 
\section{Intensity calibration of DKIST/ViSP data}\label{sec:app1}

To properly compare ViSP spectroscopic data to RHD simulations, we calibrate the ViSP spectra to absolute intensity values. To this end, we create an average ``non-flare" ViSP spectrum to compare with the Neckel solar atlas \citep{neckel1984,neckel1999}, correcting for the appropriate viewing angle (see below). The ViSP observations beginning at 22:34 UT on 2022 August 19 (yellow shaded region in Figure \ref{fig:goes_summary}(a)) best represent the quiet sun for this dataset.  

We first spatially and temporally average the quiet-Sun (QS) spectra for a selection of 500 spatial pixels and 26 time steps, selected as far as possible from the sunspot, plage, or any flaring activity. We then derive an accurate wavelength scale for this non-flare ViSP spectrum, as the wavelength values provided in the header of Level 1 (L1) DKIST data still suffer from uncertainties.\footnote{DKIST data center caveats: \url{https://nso.atlassian.net/servicedesk/customer/portal/8/article/1959985377}}

We identify two absorption lines in the averaged non-flare spectrum nominally at 396.743 and 396.926 nm, and compare the wavelength coordinates of these lines to the analogous lines in the Neckel solar atlas.\footnote{Although the chosen absorption lines would ideally be telluric, no such lines are present in the 396.8 nm window.} We adjusted the dispersion value using these reference absorption lines to find a good match between the atlas and our quiet sun profile, and produce the corrected spectral range $\lambda^*$.

Using the corrected spectral range $\lambda^*$, we then determine the multiplicative polynomial required to convert the arbitrary DN units in the L1 data to intensity ($\mathrm{\text{erg}\; \text{s}^{-1}\;\text{cm}^{-2}\; \text{sr}^{-1}\; \text{\AA}^{-1}}$) as reported in the atlas. The factor is not constant across the spectral range.  We empirically identify several wavelengths corresponding to the pseudo-continuum, and create arrays of the intensities at these wavelengths for the DKIST observations ($I_{\text{DKIST},i}$) and the solar atlas ($I_{\text{Atlas},i}$).  The multiplicative factors $\alpha_i$ used in the intensity calibrations are then:\newline
\begin{equation}\label{eq:appeq1}
    \alpha_i = I_{\text{Atlas},i} \left(\frac{I_{\text{DKIST},i}}{\psi_{\text{QS}}}\right)^{-1}
\end{equation}\label{A3}\\
where $\psi_{QS}$ is the limb darkening coefficient used to account for the center-to-limb variation in intensity.  

We then fit a second-order polynomial to the $\alpha_i$ factors to produce an array $\beta(\lambda^*)$ of multiplicative factors for each wavelength in the spectral range of the DKIST/ViSP window using \texttt{np.polyfit}.  The intensity-calibrated quiet sun and flare-time intensities $I^*_{QS,flare}(\lambda^*)$ are finally:

\begin{equation}\label{eq:appeq2}
    I^*_{\text{QS,flare}}(\lambda^*) = \beta(\lambda^*) \left( \frac{I_{\text{QS,flare}}(\lambda^*)}{\psi_{\text{QS,flare}}}\right)
\end{equation}
\newline 
\noindent where $I_{QS,flare}(\lambda^*)$ is the uncalibrated DKIST/ViSP intensity and $\psi_{QS,flare}$ is the limb darkening correction factor during the quiet sun observations and flare-time observations, respectively. 

From the Neckel limb-darkening solar spectrum \citep{neckel1984,neckel1999}, we derive $\psi_{flare} =$ 0.568 and $\psi_{QS}$ = 0.564 for this part of the solar spectrum given the position of the active region while observing.  It is necessary to include the limb darkening correction in both Equations \ref{eq:appeq1} and \ref{eq:appeq2} in case the quiet-Sun and flare-time observations are taken at different locations on the Sun.  In this instance they differ only by $\Delta\psi \sim$ 0.004, but this is not always the case.  During DKIST flare experiments performed since observing cycle 1, quiet-Sun spectra are collected as near disk center as is possible, regardless of the flare location.

For further precision, we again perform the same intensity calibration on the corrected intensity and wavelength arrays after convolving the solar atlas spectrum with an instrument point-spread-function (PSF) of full width at half maximum $w_{\text{PSF}}$, empirically determined to be 0.005 nm for these observations.

\section{Co-alignment of DKIST/ViSP and VBI with SDO}\label{sec:app2}

We use DKIST/VBI images to determine the spatial coordinates of the flare spectra from ViSP.  The 2022 August observations include images in the blue continuum at 450.3 nm (VBI-blue channel), H$\alpha$ at 656.3 nm, and TiO at 705.8 nm (both in the VBI-red channel).  

We first co-align the ViSP raster scan images with VBI to provide spatial context to ViSP spectral analysis.  Since \ion{Ca}{II}~H 396.8 nm emission corresponds to the flare ribbon location, in principle we can generally co-align the regions of \ion{Ca}{II}~H emission with the H$\alpha$ images from DKIST/VBI. Note that although both lines originate largely in the chromosphere, their source functions are not identical, and therefore the lines probe slightly different heights. For the purpose of producing approximate co-ordinates for DKIST data, we consider this approach to be sufficient.  However, for detailed comparison of the spatial differences in Balmer series and \ion{Ca}{II}~H line formation in the future, we will treat the spatial distribution of each emission line separately.  

We prepare the ViSP data for co-alignment, using the first scan as an example (Figure \ref{appendiximage}(b)).  We empirically determine the red and blue limits of the \ion{Ca}{II}~H line and integrate the line intensity along the slit for each of the four raster steps in the first scan, beginning at 20:42:07 UT. We combine the four arrays into one image.  Note that consecutive raster steps are separated in time by ${\sim}$1.8 seconds.  While the flaring chromosphere evolves appreciably over the course of a ${\sim}$7.2-second scan,  the decay phase evolution is slower than would be observed in the impulsive phase.  For co-alignment with VBI, we assume that each scan from ViSP can be treated as the same image. 

\begin{figure}[ht!]\vspace*{0.6cm}
\begin{center}
\includegraphics[width=\linewidth]{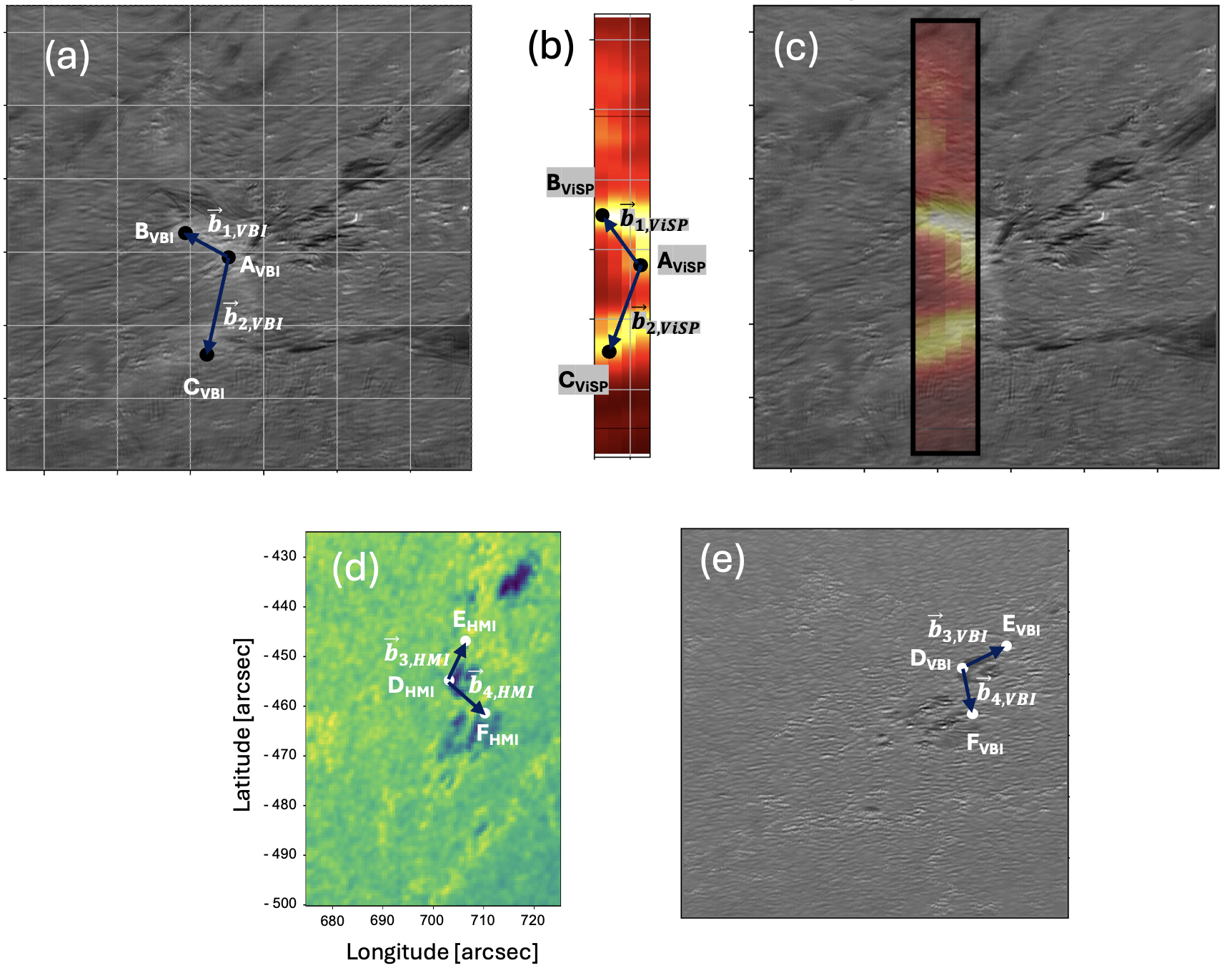}
\end{center}
\caption{Co-alignment of the ViSP and VBI images  using SDO's Helioseismic and Magnetic Imager (HMI) as a basis. (a) The VBI H$\alpha$ image frame at 2022 August 19, 20:42:07UT.  (b) The integrated \ion{Ca}{II}~H intensity from the first scan of the ViSP beginning at 20:42:07UT and ending at 20:42:13UT. In panels (a) and (b), we note the locations of the features used to co-align the ViSP and VBI images, $A_{\text{ViSP/VBI}}, B_{\text{ViSP/
VBI}}, C_{\text{ViSP/VBI}}$, and the basis vectors $\vec{b}_{1,\text{ViSP/VBI}}, \vec{b}_{2,\text{ViSP/VBI}}$ used to construct the transformation matrix.  (c) The result of the co-alignment process, showing the overlay of common chromospheric features observed with both ViSP \ion{Ca}{II}~H and VBI H$\alpha$, with both images in the VBI coordinate system. The width in each slit position of the ViSP is enlarged to clearly show ribbon structure relative to the VBI. (d) The SDO/HMI continuum image at 2022 August 19 20:42:30 UT.  (e) The DKIST/VBI TiO image from the VBI-red channel at 20:42:07UT. 
 In panels (d) and (e), we show the sample points $D_{\text{VBI/HMI}}$, $E_{\text{VBI/HMI}}$, $F_{\text{VBI/HMI}}$ and basis vectors $\vec{b}_{3,\text{VBI/HMI}}$ and $\vec{b}_{4,\text{VBI/HMI}}$ used to perform the VBI to HMI matrix transformation. }
\label{appendiximage}
\end{figure}

We then co-align the integrated ViSP \ion{Ca}{II}~H intensity images with the VBI H$\alpha$ image from the same timestep (Figure \ref{appendiximage}(a)).  We select three points ($A_{\text{ViSP/VBI}}, B_{\text{ViSP/
VBI}}$, $C_{\text{ViSP/VBI}}$) in each image, corresponding to the same three structures in both images, and define the basis vectors $\vec{b}_{1,\text{ViSP/VBI}},\; \vec{b}_{2,\text{ViSP/VBI}}$ for each coordinate space as:\newline
\begin{equation}
    \vec{b}_{1,\text{ViSP/VBI}} = B_{\text{ViSP/VBI}} - A_{\text{ViSP/VBI}}
\end{equation}

\begin{equation}
    \vec{b}_{2,\text{ViSP/VBI}} = C_{\text{ViSP/VBI}} - A_{\text{ViSP/VBI}}
\end{equation}\\
Using the basis vectors, we determine the conversion matrix $X_{\text{ViSP} \rightarrow \text{VBI}}$ that transforms the ViSP coordinate space to VBI coordinate space, defined as:\newline
\begin{equation}
    X_{\text{ViSP} \rightarrow \text{VBI}} = [\vec{b}_{1,\text{VBI}}\; \vec{b}_{2,\text{VBI}}][\vec{b}_{1,\text{ViSP}}\; \vec{b}_{2,\text{ViSP}}]^{-1}
\end{equation}\\
The new ViSP coordinates are then determined by multiplying the coordinates of each point in the ViSP image frame, relative to $A_{\text{ViSP}}$ at the origin, by $X$, and adding $A_{\text{VBI}}$, the origin of the VBI coordinate system.  The result of the co-alignment between ViSP and VBI is shown in Figure \ref{appendiximage}(c).

With VBI and ViSP images now co-aligned according to the same coordinate system from the DKIST L1 header, we determine the required rotation of the VBI co-ordinate system.  We co-align VBI and SDO images. The SDO/AIA data for chromospheric lines particularly during a flare are affected by both resolution and saturation limits.  Therefore, to co-align VBI with SDO, we use continuum images from SDO/HMI and compare them to the TiO image from VBI-red.  Figures \ref{appendiximage}(d) and (e) show the HMI and VBI data used for coalignment, respectively.  To perform the transformation from VBI to SDO space, we use the same method as described above for the co-alignment between VISP and VBI, using the points $D_{\text{VBI/HMI}}$, $ E_{\text{VBI/HMI}}$, and $F_{\text{VBI/HMI}}$ and basis vectors $b_{3,\text{VBI/HMI}}$ and $b_{4,\text{VBI/HMI}}$ to produce the conversion matrix $X_{\text{VBI} \rightarrow \text{HMI}}$.  Finally, we use this matrix to transform the VBI image points (as well as the ViSP image points, already transformed into the VBI basis) and produce the correct solar coordinates for both ViSP and VBI.  Finally, we require small manual adjustments to establish the correct VBI field-of-view.  For this step, we identify common features (such as the sunspot at point $D_{\text{HMI/VBI}}$, visible in both VBI/TiO and SDO/HMI and adjust the orientation of the VBI image to agree with SDO. Figure \ref{context}(a) shows the result.  

There are limitations to our approach for co-aligning DKIST data with SDO.  First, our method hinges on the accurate selection of common features observed by the instruments.  The spatial resolution of SDO data and the small number of raster steps in the ViSP data make accurate identification of feature position subject to significant error.  Construction of transformation matrices between the coordinate systems of the three instruments therefore requires careful oversight.  Second, we have assumed that all VBI filters are relatively co-aligned with one another. Although we may expect some differences in co-ordinates between the H$\alpha$ and TiO filters, comparison of features common to both filters (not shown) indicates that any such differences are minimal.  For our purposes we can therefore treat the H$\alpha$ and TiO images from DKIST to be sufficiently co-aligned.  This is expected, as atmospheric dispersion should be small since the two filters are relatively close in wavelength. Third, the transformation matrix between DKIST and SDO data (or even between ViSP and VBI data) is not necessarily the same for each time step, and the process must be repeated for each image frame or ViSP scan. 

\section{Comparison of RADYN and RH Ca II H line profiles}\label{sec:app3}

The impact of the statistical equilibrium assumption on the Ca II H line deserves a more thorough study outside the scope of this work.  To introduce this subject, in Figure \ref{fig:CaIIHcomp} we compare a modeled Ca II H profile from a RADYN (NEQ) simulation outside the F-CHROMA grid (beam parameters $F_{beam}=5\times10^{10}\;\mathrm{erg\;cm^{-2}\;s^{-1}}$, $\delta=8$, and $E_c=15\;\mathrm{keV}$ during a 30~s injection of energy followed by a 30~s relaxation) to modeled Ca II H from RH (assuming SE) produced with an input atmosphere from the same RADYN model.  In both cases, we impose the assumption of complete frequency redistribution (CRD).  The RH profile is generated after reading in a RADYN atmosphere as described in Section \ref{sec:RHdescrip}.  At this time step in the simulation, the line width of Ca II H is essentially the same for both cases.  The minor difference in the line core intensity does not impact our analysis of line width.  There are small differences in the Lorentzian wings, where the updated treatment of the quadratic Stark-Lo Surdo effect in RH is more appropriate \citep{zhu2019}. When considered together with the implementation of partial frequency distribution (PRD) in RH, this suggests that the RADYN+RH simulation is preferable to using RADYN alone despite the assumption of SE in RH.

\begin{figure}
\includegraphics[width=.5\linewidth]{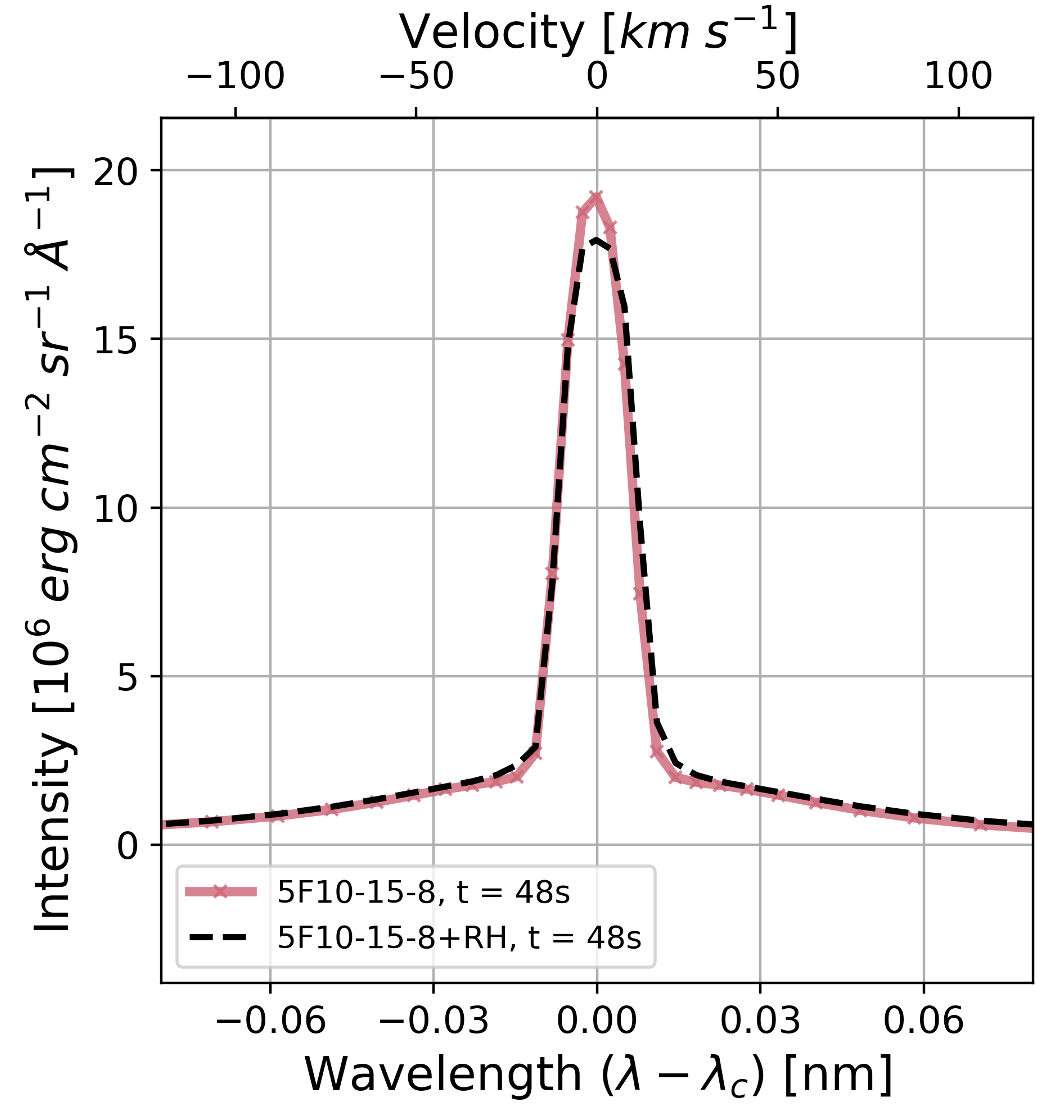}
\centering
\caption{\tiny A comparison of the RADYN (red curve) and RADYN+RH (black dotted curve) modeled profiles for the Ca II H 396.8 nm line. The simulation has a constant $\Delta t=$ 30~s injection of energy with beam parameters $F=5\times10^{10}\;\mathrm{erg\;cm^{-1}\;s^{-1}}$, $\delta=8$, and $E_c=15\;\mathrm{keV}$ followed by a 30~s relaxation phase with no energy injection.  The line profiles are shown at $t=48$~s, well into the simulation's relaxation phase.  The RADYN profile is resampled to match the spectral sampling of RADYN+RH.}
\label{fig:CaIIHcomp}
\end{figure}

%placing all citations used in bib.bib!
%\bibliography{bib,full_lib_thes_cp}{}
\bibliography{bib}
\bibliographystyle{aasjournal}

\end{document}